\documentclass[aps,prd,twocolumn,natbib,showpacs,showkeys,%
  nofootinbib]{revtex4-1}
\newif\ifprd \prdtrue           
\usepackage{ifpdf}
\usepackage[%
  \ifpdf\else hypertex,\fi
  backref,citecolor=blue,%
  pdftitle={Shadow Hamiltonians, Poisson Brackets, and Gauge Theories},%
  pdfauthor={A. D. Kennedy and Paulo Silva},%
  pdfkeywords={Hamiltonian, Poisson, Gauge Theory, Symplectic, Lattice},%
  pdfstartview={FitH},%
  pdfpagemode={FullScreen}]{hyperref}
\usepackage{bm}
\usepackage{bbm}
\begin{document}
%
%
%
\def\bb{\mathbbm}
%
%
\newif\ifdragt \dragtfalse
%
%
\newif\ifdraft \draftfalse      
\def\note[#1]#2{\message{(#1)}\ifdraft{\noindent\em[#2]\/}\fi}
%
%
\ifpdf\else\ifdraft
\fi\fi
%
%
\def\Tr{\mathop{\rm tr}}	
\def\implies{\Rightarrow}	
%
%
\def\rational#1#2{{\mathchoice{\textstyle{#1\over#2}}%
  {\scriptstyle{#1\over#2}}{\scriptscriptstyle{#1\over#2}}{#1/#2}}}
\def\half{\rational12}		
\def\third{\rational13}		
\def\quarter{\rational14}	
%
%
\def\N{{\bb N}}			
\def\Q{{\bb Q}}			
\def\R{{\bb R}}			
\def\Z{{\bb Z}}			
\def\C{{\bb C}}			
\def\ident{{\bb I}}		
%
%
\def\eqref#1{(\ref{#1})}        
\def\secref#1{\S\ref{#1}}       
\def\appref#1{Appendix~\ref{#1}} 
%
%
\def\shadow#1{{\tilde #1}}      
\def\O{{\cal O}}		
\def\vec#1{{\bm #1}}            
\def\form#1{{\bm #1}}           
\def\vp#1{{\cal #1}}		
\def\ftf{{\form\omega}}		
\def\defn{\equiv}		
\def\M{{\cal M}}		
\def\Mop{{\cal M}}              
\def\U{{\cal U}}		
\def\I{{\cal I}}                
\def\G{{\cal G}}		
\def\L{{\cal L}}		
\def\d{{\form d}}		
\def\dq{\d q}			
\def\Sym{{\cal S}}		
\def\Ham{\mathop{\rm Ham}}	
\def\Gl{\mathop{\rm Gl}}	
\def\SU{\mathop{\rm SU}}	
\def\su{\mathop{\rm su}}	
\def\Aut{\mathop{\rm Aut}}	
\def\dt{\delta\tau}		
\def\CT#1#2{[#1,#2]}		
\def\CK#1#2{\left\langle#1,#2\right\rangle} 
\def\ad{\mathop {\rm ad}}	
\def\pad{\mathop {\widehat{\rm ad}}} 
\def\Re{\mathop {\rm Re}}	
\def\Im{\mathop {\rm Im}}	
\def\AHP{\mathop {\cal A}}	
\def\SHP{\mathop {\cal H}}	
\def\TAHP{\mathop {\cal T}}	
\def\SPF{S_{\mbox{\tiny PF}}}	
\def\plv{basic lattice vector}  
\def\Plv{Basic Lattice Vector}  
\ifdragt
  \def\colonop#1{\mathopen:#1\mathclose:}
  \def\HV#1{\colonop{#1}}	
  \def\PB#1#2{[\colonop{#1,#2}]} 
\else
  \def\HV#1{{\setbox0=\hbox{$#1$}%
     \ifdim\wd0>1em \widehat\vec{{#1}}\else\hat\vec{{#1}}\fi}}
  \def\PB#1#2{\{#1,#2\}}
\fi
%
%
\def\sqr#1#2{{\vcenter{\hrule height.#2pt
   \hbox{\vrule width.#2pt height#1pt \kern#1pt
      \vrule width.#2pt}
   \hrule height.#2pt}}}
\def\stpl#1#2{{\vcenter{\hrule height.#2pt
   \hbox{\vrule width.#2pt height#1pt \kern#1pt
      \vrule width.#2pt}}}}
\def\plaq{{\mathop{\,\mathchoice\sqr64\sqr64\sqr{4.2}3\sqr{3.0}3\,}}}
\def\staple{{\mathop{\,\mathchoice\stpl64\stpl64\stpl{4.2}3\stpl{3.0}3\,}}}
%
%
\def\dd#1#2{{\mathchoice{d#1\over d#2}%
  {d#1\!/\!d#2}{d#1\!/\!d#2}{d#1\!/\!d#2}}}
\def\pdd#1#2{{\mathchoice{\partial#1\over\partial#2}%
  {\partial#1\!/\!\partial#2}%
  {\partial#1\!/\!\partial#2}%
  {\partial#1\!/\!\partial#2}}}
\def\pddsq#1#2{{\mathchoice{\partial^2#1\over\partial{#2}^2}%
  {\partial^2#1\!/\!\partial{#2}^2}%
  {\partial^2#1\!/\!\partial{#2}^2}%
  {\partial#1^2\!/\!\partial{#2}^2}}}
%
%
\hyphenation{pseudo-fermion}
\title{Shadow Hamiltonians, Poisson Brackets,\\ and Gauge Theories}
\ifprd
\author{A.~D.~Kennedy}
\affiliation{School of Physics \& Astronomy and SUPA,
  The University of Edinburgh, The King's Buildings,
  Edinburgh, EH9~3JZ, Scotland}
\author{P.~J. Silva}
\affiliation{Centro de F{\'\i}sica Computacional,
  Departamento de F{\'\i}sica,
  Universidade de Coimbra,
  Rua Larga, Coimbra 3004-516, Portugal}
\author{M.~A.~Clark}
\affiliation{NVIDIA Corporation,
  2701 San Tomas Expressway,
  Santa Clara, CA 95050, USA}
\date{\today}
\fi
\begin{abstract}
  \noindent Numerical lattice gauge theory computations to generate gauge field
  configurations including the effects of dynamical fermions are usually
  carried out using algorithms that require the molecular dynamics evolution of
  gauge fields using symplectic integrators.  Sophisticated integrators are in
  common use but are hard to optimise, and force-gradient integrators show
  promise especially for large lattice volumes.  We explain why symplectic
  integrators lead to very efficient Monte Carlo algorithms because they
  exactly conserve a shadow Hamiltonian.  The shadow Hamiltonian may be
  expanded in terms of Poisson brackets, and can be used to optimize the
  integrators. We show how this may be done for gauge theories by extending the
  formulation of Hamiltonian mechanics on Lie groups to include Poisson
  brackets and shadows, and by giving a general method for the practical
  computation of forces, force-gradients, and Poisson brackets for gauge
  theories. {\parfillskip=0pt\par}
\end{abstract}
\ifprd
\maketitle
\fi

\section{Introduction}

Essentially all algorithms used in lattice gauge theory computations to
generate gauge field configurations including the effects of dynamical fermions
are variants of the Hybrid Monte Carlo (HMC) algorithm \cite{duane87a}, which
requires a reversible and area-preserving integrator for its molecular dynamics
step.  The simplest such method is the leapfrog integrator, but there is a
large class of \emph{symplectic integrators} \cite{hairer:2006} that have these
properties and are potentially more cost-effective.  Indeed, many
state-of-the-art computations use the second order minimum norm integrator
\cite{omelyan:2001, Takaishi:1999bi, forcrand:2006} which has a free parameter,
which heretofore has been tuned in an \emph{ad hoc} manner.

The formulation of Hamiltonian dynamics on Lie group manifolds, which is
required for molecular dynamics on gauge fields \cite{drummond83a, kennedy88b},
and the fact that symplectic integrators conserve a shadow Hamiltonian are well
known; our goal is to combine the two and show how to construct the shadow
Hamiltonian for gauge theories.  This is most easily done using the formalism
of differential forms \cite{bishop:1980, bruhat:1977, helgason:1978,
  hicks:1971, spivak:1970}; in order to fix our notation and establish the
necessary results, some of which are not easy to find in the literature, we
provide a brief review in \appref{sec:diff-forms}.

The shadow Hamiltonian is expressed as an asymptotic expansion in the
integration step size \(\dt\) whose coefficients depend on the parameters
specifying the integrator under consideration and a collection of Poisson
brackets.  These Poisson brackets are complicated functions on phase space,
where in the case of gauge field molecular dynamics a point in phase space is
an entire gauge field configuration and its associated ``fictitious'' momenta.
For extensive systems such as field theories, unlike the few body systems
considered previously \cite{omelyan:2002, omelyan:2003}, the values of the
Poisson brackets have a distribution that is sharply peaked about their mean
values when we choose the starting points of their molecular dynamics
trajectories to be chosen from the distribution \(e^{-H}\), as is done in the
HMC algorithm.  This may be understood as a consequence of the central limit
theorem applied to the contributions to the Poisson brackets coming from many
independent regions of space-time.  This means that for configurations that
occur with non vanishingly small probability the shadow Hamiltonian may be
considered to be a function of the average values of the Poisson brackets; if
these are measured on a few test trajectories then the integrator parameters
may be chosen to minimize the computational cost \cite{Clark:2008gh,
  Clark:2010qw, Clark:2011ir}.  Perhaps surprisingly this does not correspond
to minimizing the average difference between the Hamiltonian and its shadow
\footnote{Since the shadow is only defined up to an additive constant this
  cannot be too surprising.}, and instead to minimizing the variance of the
distribution of the shadow.  We shall not be concerned with the details of this
tuning procedure here, but we refer the interested reader to
\cite{Clark:2011ir} for details: instead, the aim of this paper is to explain
how the Poisson brackets, forces, and force-gradients may be computed at any
given point in phase space.

In \cite{kennedy88b} expressions for the molecular dynamics force were derived
from the classical mechanics specified by the Hamiltonian function and a
suitable chosen group-invariant fundamential two-form.  We extend this analysis
to obtain an expression for the force-gradient for gauge fields
\cite{Kennedy:2009fe}, which can be used to provide a ``second derivative''
integrator step for the construction of improved integrators
\cite{omelyan:2002, omelyan:2003}.

\subsection{Multiple link updates} \label{sec:multiple-links}

For much of this paper we shall be considering a Hamiltonian system with a
phase space which is the cotangent bundle \(T^{*}\G\) over a base space that is
a Lie group manifold \(\G\) and whose fibres are isomorphic to its Lie algebra.
We shall call the cotangent one-forms ``momenta'', although in the context of
HMC they are called ``fictitious momenta'' as they are quite different from the
canonical momenta of the underlying field theory.  For a gauge field theory we
may associate such a phase space with every link of the lattice.  One might at
first think that we need to introduce some fibre bundle structure over the
space-time lattice itself, but fortunately that is not necessary.  We can
consider the molecular dynamics evolution of each gauge link separately; they
are coupled together through the action that plays the r\^ole of the potential
energy part of the Hamiltonian, but the kinetic energy part does not couple
different links.  For HMC we are free to choose the form of the kinetic energy,
so we can take it to be of the form \footnote{For notational simplicity we
  consider here a theory with a scalar field \(\phi\) and the corresponding
  momentum \(p\) defined on the links of a lattice.} \(T(p) = \half\sum_\ell
c_\ell p_\ell^2\) where \(p_\ell\) is the momentum associated with the link
\(\ell\), and \(c_\ell\) is a link-dependent coefficient that is constant in
molecular dynamics ``fictitious'' time.  If we wish to evolve the single link
\(\phi_\ell\) on its own we can choose \(c_{\ell'} = \delta_{\ell, \ell'}\) so
that \(\dot\phi_{\ell'} = \pdd H{p_{\ell'}} = \pdd T{p_\ell} =
c_{\ell'}p_{\ell'} = 0\) if \(\ell\neq\ell'\).  We are also free to choose
\(c_\ell=1\) for all links, which is the usual situation where we update the
gauge field simultaneously across the entire lattice.  Another interesting
choice for the kinetic energy is to choose \(c_\ell=1\) for all spatial links
and \(c_\ell=\xi\) for all temporal ones: this is the procedure suggested in
\cite{Morrin:2006tf, Lin:2008pr} for evolving anisotropic lattices \footnote{In
  \cite{Lin:2008pr} the temporal step size is adjusted rather than the kinetic
  energy, but this is equivalent after a rescaling of the temporal momenta.}.
The momentum anisotropy \(\xi\) is a parameter that can be adjusted to optimize
the HMC algorithm for a given anisotropy in the action; if the spatial and
temporal contributions to the Poisson brackets are measured separately then the
techniques of \cite{Clark:2011ir} can be used to tune~\(\xi\) along with other
integrator parameters.

\subsection{Pseudofermion forces} \label{sec:pf-forces}

So far we have only been discussing pure gauge theories, but in practice the
cost of most lattice computations is dominated by the inclusion of fermions.
This is because we need to solve a large system of linear equations in order to
update the fictitious moments (i.e., to apply the Hamiltonian vector field
\(\HV S\) in the notation we will introduce later).  Typically we have an
action \(S\) which is the sum of a pure gauge part \(S_G\), built out of sums
of small Wilson loops (traces of a closed loops of gauge links) such as
plaquettes, and a pseudofermion part \(S_F\) built out of sums of pairs of
pseudofermion fields \(\phi\) connected by a string of gauge links.  If we want
to compute the force acting on a particular gauge link \footnote{We shall refer
both to a gauge link variable and the link on which it lives as \(U\) when
there is no ambiguity.}  \(U\) then it is convenient to write \(S_G=\Re
\Tr(\staple U)\) and \(S_F=\phi^\dagger\Mop^{-1}(U)\phi\) where the ``staple''
\(\staple\) is the sum of all gauge link strings that connect the ends of the
link \(U\) that correspond to the Wilson loops in \(S_G\), and the Hermitian
lattice matrix \(\Mop(U)\) is the sum of all gauge link strings that include
\(U\) that occur in~\(S_F\).  For a local action all of these strings are in
some neighbourhood of \(U\), and we have dropped all other terms in the action
because they are independent of \(U\) and therefore do not contribute to the
force on that particular link.  In reality we update many or all the links on
the lattice at once, so we compute the force on each link in parallel.  By the
``force'' we mean the quantity \(\vec e_i(S)T^i\) where \(\vec e_i\) is a
linear differential operator (vector field) whose action on \(U\) is specified
by \(\vec e_i(U)=-T_iU\) and which we shall define carefully
later~\eqref{eq:eiu}, and \(T^i\) is the representation of a generator of the
gauge group.  It is important to note that here \(\vec e_i\) acts only on the
gauge link \(U\), it gives zero if applied to any other link variable.  There
is an opportunity for confusion when we refer to \(\vec e_i\) as a vector
field; it is a vector field defined over the phase space of the link \(U\), but
it is \emph{not} a field over the space-time lattice.  In order to reduce
confusion we refer to quantities defined over the space-time lattice as
\emph{lattice vectors}, and space time linear differential operators such as
the Dirac operator (or more precisely lattice difference operators acting on
lattice vectors such as the Wilson--Dirac operator) as \emph{lattice matrices}.

The contribution to the force from the pure gauge part of the action is \(\vec
e_i(S_G)T^i=\Re\Tr\Bigl(\staple \vec e_i(U)\Bigr)T^i=-\Re\Tr(\staple T_iU)T^i =
-\Re\Tr(U \staple T_i)T^i=-a\TAHP(U \staple)\), \(\TAHP\) being the projector
onto the Lie algebra, that is \(\TAHP(X)= \Re\Tr(XT_i)T^i/a\) where there is an
implicit sum over \(i\) as usual and the generators \(T_i\) are normalized such
that \(\Tr(T_iT_j)=a\delta_{ij}\).  If the gauge group is \(\SU(N)\) and we
choose its generators to be anti-Hermitian so as not to introduce artificial
factors of \(i\), then \(\TAHP(X)\) is just the traceless anti-Hermitian part
of~\(X\).

The pseudofermion contribution to the force is \(\vec e_i(S_F)T^i=
\phi^\dagger\vec e_i\Bigl(\Mop(U)^{-1}\Bigr)\phi T^i\).  Since \(\vec e_i\) is
a linear differential operator we have \(0=\vec e_i(\ident)= \vec e_i(\Mop
\Mop^{-1})=\vec e_i(\Mop)\Mop^{-1}+\Mop\vec e_i(\Mop^{-1})\), and hence \(\vec
e_i(\Mop^{-1})=-\Mop^{-1} \vec e_i(\Mop)\Mop^{-1}\).  Therefore \(\vec e_i(S_F)
T^i=-\Re\Tr\left[\vec e_i\Bigl(\Mop(U)\Bigr)X\otimes X^\dagger\right]T^i\),
where we have defined \(X \defn\Mop^{-1}\phi\) to be the solution of a large
but sparse system of linear equations (since \(\Mop\) is local), this may be
computed on all lattice sites and used to update some or all gauge links in
parallel.  The outer product \(X\otimes X^\dagger\) is the rank one Hermitian
lattice matrix whose action on an arbitrary lattice vector \(y\) is
proportional to the projection of \(y\) along \(X\), namely \((X\otimes
X^\dagger)y= X(X^\dagger y)\).

We can express the pseudofermion action in the form \(S_F=-\Re\Tr[\Mop(U)X
\otimes X^\dagger]\) analogous to that of \(S_\G\) if we consider \(X\) to be
a lattice vector that is independent of~\(U\).  This means that once we have
computed \(X\) the calculation of the gauge and pseudofermion parts of the
force and related quantities are very similar.  Both the gauge and
pseudofermion actions can be written as the trace of lattice operators times
\(U\), where the lattice operators are either local (\(\staple\) and \(\Mop\))
or low rank (\(X\otimes X^\dagger\)).  Both local and low rank operators are
relatively cheap to apply to lattice vectors or to trace, the former only
involving links in the neighbourhood of \(U\), and the latter only involving
inner products of lattice vectors.  For example, we may evaluate the trace
\(\Tr[\Mop(U)X\otimes X^\dagger]=X^\dagger\Mop(U)X\) as the inner product of
\(X^\dagger\) with the vector \(\Mop(U)X\).

If we include spin degrees of freedom then we must replace \(X\otimes
X^\dagger\) by a sum of outer products for each spin component, but the result
is still a low rank matrix which is therefore cheap to apply.  Likewise if we
wish to introduce \(n\) pseudofermion fields so as to reduce the noise in the
stochastic estimate of the fermionic force and thus defer the breakdown in the
asymptotic expansion for the shadow Hamiltonian to significantly larger
integrator step sizes \cite{Hasenbusch:2001ne, Hasenbusch:2002ai,
  Luscher:2005rx, Clark:2006fx, Clark:2006wp}, then we only increase the rank
by a factor of \(n\).

\subsection{Outline}

The structure of this paper is as follows.  In \secref{sec:ham-mech} we
consider the general formulation of Hamiltonian mechanics on a symplectic
manifold \cite{berndt:2001}; this serves to introduce the important concepts of
the fundamental 2-form, the Hamiltonian vector field it associates with any
0-form, and the Poisson bracket of two 0-forms.  We show that Poisson brackets
satisfy the Jacobi identity, and that the commutator of two Hamiltonian vector
fields is itself a Hamiltonian vector field, and explain the isomorphism
between the Lie algebra of commutators of Hamiltonian vector fields and that of
Poisson brackets of 0-forms.  The reason we need all this mathematical
machinery is that when we consider Hamiltonian mechanics on Lie groups in
\secref{sec:ham-lie} we will introduce a non-trivial fundamental 2-form in
order to make the dynamics symmetric under the action of the group.  Moreover,
the fact that Hamiltonian vector fields form a Lie algebra is crucial for the
definition of the shadow Hamiltonian, which we give in~\secref{sec:symp-int}.
The exposition assumes some knowledge of the theory of differential forms, an
overview of which is given in \appref{sec:diff-forms}.

\secref{sec:symp-int} introduces symplectic integrators by noting that if a
0-form on phase space only depends on the momenta \(p\) or only on the
positions \(q\) then the integral curves of its Hamiltonian vector field are
easily found.  We are interested in Hamiltonians \(H(q,p)=T(p)+S(q)\) that are
the sum of two such functions, and we show how this allows us to construct
symplectic integrators to find approximate integral curves for \(\HV H\) using
the Baker--Campbell--Hausdorff (BCH) formula.  We give some simple examples of
integrators for a system on a symplectic manifold with fundamental 2-form
\(\ftf = \d q\wedge\d p\), and show how to compute the corresponding shadow
Hamiltonians.  When the kinetic energy is of the form \(T(p) = \half p^2\) we
show that the Poisson bracket \(\PB S{\PB ST}\) is independent of \(p\) and
explain how it may thus be used to construct a force-gradient integrator step.

\secref{sec:ham-lie} defines a symplectic structure on Lie group manifolds, or
more precisely on their cotangent bundle \(T^{*}\G\), that is compatible with
the group structure.  This is done by introducing the natural fundamental
2-form terms of Maurer--Cartan forms, and it is here that the mathematical
framework we have developed becomes necessary.  We derive explicit
formul\ae\ for Hamiltonian vector fields and Poisson brackets in terms of the
momentum coordinates (which are well-defined globally) and the family of
left-invariant vector fields dual to the Maurer--Cartan forms.  All the
independent Poisson brackets of \(S\) and \(T\) that can occur in shadow
Hamiltonians up to and including \(\O(\dt^4)\) are given explicitly for the
case where \(S\) is momentum-independent and \(T\) is quadratic in the momenta.
We then show how to express the results in terms of matrix representations of
the Lie group, as these are what is used in practice.

In \secref{sec:SUN} we evaluate the formul\ae\ for the Poisson brackets for the
physically interesting case of the fundamental representation of~\(\SU(N)\).
We show that they can all be expressed as traces of a collection of
Lie-algebra-valued quantities: as these live on links we name them {\plv}s.

In \secref{sec:towers} we address the problem of computing these {\plv}s. We do
this first for the simple case where only a single link is updated, and then
introduce the algebra of \emph{towers} to give an efficient way of computing
them in general.

\appref{sec:diff-forms} gives a brief survey of the theory of differential
forms and serves to fix our notation and conventions, as does
\appref{sec:lie-groups} which gives an overview of the properties of Lie
groups.

\section{Hamiltonian Mechanics} \label{sec:ham-mech}

\subsection{Symplectic Manifolds}

A Hamiltonian system is defined on \emph{phase space} which is a differential
manifold \(\M\) with a \emph{symplectic structure} given by some
\emph{fundamental 2-form} \(\ftf\) that is closed, \(\d\ftf=0\), and globally
invertible.  Phase space is usually the cotangent bundle \(T^{*}\G\) over some
\emph{configuration space} manifold \(\G\).  For every 0-form \(F\in
\Lambda^0\) on \(\M\), that is for every \(C^\infty\) smooth function \(F:
\M\to\M\), there is a corresponding \emph{Hamiltonian vector field} \(\HV F\in
\Ham\M\) such that \(\d F\defn i_{\HV F}\ftf\): in other words \(\d F(\vec y) =
(i_{\HV F}\ftf)(\vec y) = \ftf(\HV F,\vec y)\) for any vector field \(\vec y\).

A 0-form \(Z\) corresponds to a vanishing Hamiltonian vector field iff \(\d
Z=0\), so we have the following short exact sequence \(0\to\R\to\Lambda^0(\M)
\to\Ham\M\to0\).  This implies that there is a bijective diffeomorphism
\(\Lambda^{0}(\M)/\R \leftrightarrow \Ham\M\).  The nature of this
correspondence between 0-forms (up to an additive constant) and Hamiltonian
vector fields will be examined further in the following sections.

\subsection{Poisson Brackets}

Consider the action of a Hamiltonian vector field \(\HV F\) on a 0-form \(G\),
\[\HV FG = \d G(\HV F) = i_{\HV G}\ftf(\HV F) = \ftf(\HV G,\HV F) \defn \PB
FG,\] where in the first equality we have made use of the definition of the
exterior derivative of a 0-form \(G\) acting on an arbitrary vector field
\(\vec y\), \(\d G(\vec y) \defn\vec yG\), and in the last equality we have
introduced the \emph{Poisson bracket} \(\PB AB\defn-\ftf(\HV A,\HV B)\) for
any pair of 0-forms \(A\) and \(B\). The minus sign has to appear somewhere,
and our convention is to introduce it here in the definition of the Poisson
bracket.

\subsection{Jacobi Identity}

The invariant expression~\eqref{eq:d-2-form} for the exterior derivative
\(\d\ftf\) of a 2-form \(\ftf\) applied to three arbitrary vector fields \(\vec
x\), \(\vec y\), and~\(\vec z\)
\begin{eqnarray*}
  \d\ftf(\vec x,\vec y,\vec z)
  &=& \vec x\ftf(\vec y,\vec z) + \vec y\ftf(\vec z,\vec x)
    + \vec z\ftf(\vec x,\vec y) \\
  && - \ftf(\CT{\vec x}{\vec y},\vec z) - \ftf(\CT{\vec y}{\vec z},\vec x)
    - \ftf(\CT{\vec z}{\vec x},\vec y),
\end{eqnarray*}
displays an interesting cyclic symmetry in the three vector fields \(\vec x\),
\(\vec y\), and \(\vec z\).  This has an important consequence if \(\ftf\) is
the fundamental 2-form and the vector fields are Hamiltonian: if \(A\), \(B\),
and \(C\) are three arbitrary 0-forms then \[\HV A\ftf(\HV B,\HV C) = -\HV
A\PB BC = -\PB A{\PB BC},\] and also
\begin{eqnarray*}
  \ftf(\CT{\HV A}{\HV B},\HV C) 
  &=& -\ftf(\HV C,\CT{\HV A}{\HV B})
  = -\d C(\CT{\HV A}{\HV B}) \\
  &=& -\CT{\HV A}{\HV B}C
  = (\HV B\HV A - \HV A\HV B)C \\
  &=& \PB B{\PB AC} - \PB A{\PB BC}.  
\end{eqnarray*}
We thus find that the condition \(\d\ftf=0\) implies that the cyclic sum of of
nested Poisson brackets must vanish, \(\d\ftf(\HV A,\HV B,\HV C) = \PB A{\PB
BC} + \PB B{\PB CA} + \PB C{\PB AB} = 0\): this is just the \emph{Jacobi
identity} which, together with the antisymmetry of the Poisson bracket,
demonstrates that 0-forms on \(\M\) together with the product given by the
Poisson bracket form a \emph{Lie algebra}.

We can use the Jacobi identity to derive another useful result.  The commutator
of any two vector fields is a vector field (q.v.,
equation~\eqref{eq:commutator}); if both vector fields are Hamiltonian then
their commutator is also Hamiltonian, since
\begin{eqnarray*}
  \CT{\HV A}{\HV B}C &=& (\HV A\HV B-\HV B\HV A)C
  = \PB A{\PB BC} - \PB B{\PB AC} \\
  &=& -\PB C{\PB AB} = \PB{\PB AB}C = \HV{\PB AB}C  
\end{eqnarray*}
where we applied the Jacobi identity in the antepenultimate step.  Since this
must hold \(\forall C\in\Lambda^0\) we have
\begin{equation}
  \CT{\HV A}{\HV B} = \HV{\PB AB} \in\Ham\M \label{eq:Poisson-Lie}
\end{equation}
telling us that not only is the commutator of two Hamiltonian vector fields
Hamiltonian as promised, but also that it corresponds to the 0-form that is
the Poisson bracket of the 0-forms corresponding to the original pair of
Hamiltonian vector fields.  The bijection \(\Lambda^0(\M)/\R\leftrightarrow
\Ham(\M)\) is therefore an isomorphism of Lie algebras.

\subsection{Lie Derivatives and Equations of Motion}

Given a Hamiltonian \(H\in\Lambda^0(\M)\) and a fundamental 2-form \(\ftf\) we
may construct the Hamiltonian vector field \(\HV H\), and for any point
\(p\in\M\) we may --- at least locally --- define an \emph{integral curve}.  We
may also define a \emph{local flow} \(\sigma:\I\times\U\to\M\) of trajectories
starting at any point \(p\in\U \subseteq\M\) in some neighbourhood of \(p\),
\(\sigma:\R\to\M\), satisfying Hamilton's equations \(\dd\sigma t=\HV H\) and
the initial condition \(\sigma(0) = p\).  Hamilton's equations are thus most
naturally expressed in terms of Lie derivatives (\secref{sec:lie-derivatives}),
\(\dd{\vec T}t = \L_{\HV H}\vec T\), for any tensor \(\vec T\).  In particular
a scalar field (0-form) \(F\), vector field \(\vec v\), and 1-form
\(\form\theta\) must obey
\begin{eqnarray*}
  \dd Ft &=& \L_{\HV H}F = \HV HF = \PB HF, \\
  \dd{\vec v}t &=& \L_{\HV H}\vec v = \CT{\HV H}{\vec v}, \\
  \mbox{and} \qquad \dd{\form\theta}t &=& \L_{\HV H}\form\theta
    = (i_{\HV H}\d + \d i_{\HV H})\form\theta.
\end{eqnarray*}
The formal solution of the equation of motion \(\dd{\vec T}t = \L_{\HV H}\vec
T\) is \(\vec T(t) = \exp(t\L_{\HV H})\vec T(0)\), where the exponential
function is defined as \(\exp(t\L_{\HV H}) = \lim_{n\to\infty} \left(1 +
\frac tn\L_{\HV H}\right)^n = \sum_{j=0}^\infty (t\L_{\HV H})^j/j!\).

\section{Symplectic Integrators and Shadow Hamiltonians} \label{sec:symp-int}

\subsection{Baker--Campbell--Hausdorff Formula}

The BCH formula states that if \(A\) and \(B\) belong to an associative algebra
then 
\begin{equation}
  \ln(e^Ae^B) = \sum_{n=1}^\infty c_n(A,B) \label{eq:hausdorff-series}
\end{equation}
where the \(c_n\), belonging to the \emph{free Lie algebra} \footnote{That is
the Lie algebra whose Lie bracket is the commutator constructed from the
associative product.  For more details about free Lie algebras and a proof of
the BCH formula see Appendix~B of~\cite{Clark:2007ff}.}, are recursively
determined from the relations \(c_1=A+B\) and
\begin{eqnarray}
  \lefteqn{(n+1)c_{n+1} =} \qquad && \nonumber \\
  && \sum_{m=1}^{\lfloor n/2\rfloor} {B_{2m}\over(2m)!}
      \!\!\! \sum_{{k_1,\ldots,k_{2m}\geq1}\atop{k_1+\cdots+k_{2m}=n}} \!\!\!\!
      \ad c_{k_1}\ldots\ad c_{k_{2m}} (A+B) \nonumber \\ 
  && \qquad\qquad\quad - \half(\ad c_n)(A-B) \qquad\mbox{for \(n\geq1\),}
  \label{eq:hausdorff-coeffs}
\end{eqnarray}
where \(\ad a:b\mapsto\CT ab\) and the \emph{Bernouilli numbers} \(B_n\) are
defined by \[\frac x{e^x - 1} \defn \sum_{n\geq0} \frac{B_nx^n}{n!}.\]

The first few terms in the Hausdorff series are
\def\1{\scriptstyle}
\[
  \begin{array}{lr}
    \1 \ln(e^Ae^B) = (A + B) + \half[A,B] & \\[0.5ex]
    \multicolumn{2}{r}{\1 + \rational1{12} ([A,[A,B]] - [B,[A,B]])
      - \rational1{24}[B,[A,[A,B]]]} \\[1ex]
    \multicolumn{2}{r}{\1 + \rational1{720}\left( 
      \begin{array}{rr}
	\1- 4 [B,[A,[A,[A,B]]]] & \1- 6 [[A,B],[A,[A,B]]] \\
        \1+ 4 [B,[B,[A,[A,B]]]] & \1- 2 [[A,B],[B,[A,B]]] \\
        \1- [A,[A,[A,[A,B]]]] & \1+ [B,[B,[B,[A,B]]]] 
      \end{array}\right) + \cdots}
  \end{array}
\]
From this we easily obtain the corresponding formula for a symmetric product
\begin{equation}
  \begin{array}{lr}
    \1 \ln(e^{A/2}e^Be^{A/2}) = (A + B)
      - \rational1{24} (2 [B,[A,B]] + [A,[A,B]]) & \\[0.5ex]
    \multicolumn{2}{r}{\1 + \rational1{5760} \left(
      \begin{array}{rr}
        \1 32 [B,[B,[A,[A,B]]]] & \1 - 16 [[A,B],[B,[A,B]]] \\
	\1 + 28 [B,[A,[A,[A,B]]]] & \1 + 12 [[A,B],[A,[A,B]]] \\
	\1 + 8 [B,[B,[B,[A,B]]]] & \1 + 7 [A,[A,[A,[A,B]]]]
      \end{array}\right) + \cdots}
  \end{array}
  \label{eq:bchsym}
\end{equation}

\subsection{Symplectic Integrators}

The integral curve of a Hamiltonian vector field \(\HV A\) is given by the
exponential map \(t\mapsto\exp(t\HV A)\) acting on the initial point.  Given
two Hamiltonian vector fields \(\HV A\) and \(\HV B\) we can construct a curve
that is alternately tangential to each vector field from the composition of
their exponential maps \(t\mapsto[\exp(t\HV A/n)\exp(t\HV B/n)]^n\) for some
\(n\in\N\).  Such a map is called a \emph{symplectic integrator} as it
manifestly preserves the symplectic structure since each individual exponential
map does.  The BCH formula \eqref{eq:hausdorff-series} tells us that this curve
is in fact itself the integral curve of a vector field \(\vec D_{t/n}\)
\begin{eqnarray*}
  \lefteqn{\left[\exp\left(\frac{t\HV A}n\right)
    \exp\left(\frac{t\HV B}n\right)\right]^n} \qquad && \\
  &=& \left[\exp\left((\HV A+\HV B)\frac tn
    + \sum_{m=2}^\infty c_m(\HV A,\HV B)\left(\frac tn\right)^m\right)
    \right]^n \\
  &=& \exp\left[\left(\HV A+\HV B + \sum_{m=2}^\infty c_m(\HV A,\HV B)
    \left(\frac tn\right)^{m-1}\right)t\right] \\
  &=& \exp\left(\vec D_{t/n}t\right),
\end{eqnarray*}
where \(\vec D_\varepsilon \defn \HV A + \HV B+ \sum_{m=2}^\infty c_m(\HV A,\HV
B) \varepsilon^{m-1}\).  As all the \(c_m\) are commutators, equation
\eqref{eq:Poisson-Lie} tells us that \(\vec D_\varepsilon\) is a Hamiltonian
vector field corresponding to the \emph{shadow} 0-form \(D_\varepsilon\) under
the isomorphism \(\Ham\M \leftrightarrow \Lambda_0(\M)/\R\) discussed before.
In other words, \(\vec D_\varepsilon=\HV D_\varepsilon\) where the 0-form
\(D_\varepsilon \defn A+B + \sum_{m=2}^\infty c'_m(A,B)\varepsilon^{m-1}\) with
the \(c'_m\) defined by \eqref{eq:hausdorff-coeffs} in terms of the Poisson
bracket image of the adjoint under the Lie algebra isomorphism
\eqref{eq:Poisson-Lie} \(\ad\HV A\mapsto\pad A\) where \(\pad A: B\mapsto\PB
AB\).  We note in passing that the shadow is only defined up to an additive
constant.

The BCH formula is obtained by formal manipulation of the exponential series,
so we should choose a sufficiently large \(n\) to ensure that the Hausdorff
series converges.  In order to study the convergence of the BCH formula we need
to specify a topology on the space of Hamiltonian vector fields \(\Ham\M\).  It
is simpler to ask the same question about the convergence of the corresponding
expansion for the shadow Hamiltonian, for which there is an obvious topology as
the coefficients are 0-forms and we can use the usual \(L_p\) norms.  In most
cases of interest none of these norms are bounded, so the series is only
asymptotic at best.  In HMC the momenta are selected from a Gaussian
distribution \(e^{-T(p)}\), so the values of the Poisson brackets can become
arbitrarily large, but with exponentially small probability.  There is no value
of \(\varepsilon\) for which the Hausdorff series always converges, but it
might well be that for any \(\delta>0\) we can find an \(\varepsilon>0\) such
that it does converge with probability \(>1-\delta\).  This may be acceptable
for HMC, where an exponentially small chance of a trajectory becoming unstable
is unimportant: it will presumably be rejected and the next momentum or
pseudofermion refreshment will resolve the problem.  If the large norm comes
from the gauge field configuration then there could be more severe problems.

\subsection{Symmetric Symplectic Integrators}

In general a symplectic integrator is not reversible, that is the group
commutator \[\exp(-t\HV A/n)\exp(-t\HV B/n)\exp(t\HV A/n)\exp(t\HV B/n)\neq
\ident;\] indeed we immediately see from this expression that that the
integrator is reversible iff \(\CT{\HV A}{\HV B}=0\).  This blemish is easily
eradicated by using a \emph{symmetric symplectic integrator}, such as
\(\exp(t\HV A/2n)\exp(t\HV B/n)\exp(t\HV A/2n)\).  An additional advantage of
such integrators is that only even powers of \(\varepsilon\) occur in the
Hausdorff series for their shadow Hamiltonians \(D_\varepsilon\), so
\(A+B-D_\varepsilon = \O(\varepsilon^2)\), making them better approximations to
the exponential map of \(\HV A+\HV B\) itself.

\subsection{Practical Integrators}

Finding a closed-form expression for the integral curve of some Hamiltonian
vector field \(\HV A\) is impossible in most cases as there is no closed-form
solution of Hamilton's equations.  However, there are some special cases where
we can find such a solution.

For example, suppose that in some local patch of phase space with coordinates
\(q\) and \(p\) the fundamental 2-form is \footnote{We can always find
  coordinates for which this is true according to Darboux's theorem.}
\(\ftf=\d q\wedge\d p\), \(A\) is an arbitary 0-form, \(\vec X\) is an
arbitrary vector field on phase space.  Then
\begin{eqnarray*}
  \d A &=& \pdd Aq\d q + \pdd Ap\d p, \\
  \vec X &\defn& X_q\pdd{}q + X_p\pdd{}p, \\
  \HV A &\defn& A_q\pdd{}q + A_p\pdd{}p,
\end{eqnarray*}
and we have
\begin{eqnarray*} 
  \d A(\vec X) &=& \pdd AqX_q + \pdd ApX_p = \ftf(\HV A,\vec X) \\
  &=& (\d q\wedge\d p)
    \left(A_q\pdd{}q + A_p\pdd{}p,X_q\pdd{}q + X_p\pdd{}p\right) \\
  &=& A_qX_p - A_pX_q.
\end{eqnarray*}
Since \(\vec X\) is arbitary we can equate coefficients of \(X_q\) and \(X_p\)
to obtain \[\HV A = \pdd Ap\,\pdd{}q - \pdd Aq\,\pdd{}p.\] Let \(c(t) =
(q_t,p_t)\) be the integral curve of \(\HV A\) with \(c(0) = (q_0,p_0)\), which
means that for any 0-form \(f\) it must satisfy the differential
equations \[(\HV Af)\circ c = \dd{}t (f\circ c),\] or equivalently \[\dot q_t =
\pdd Ap(q_t,p_t) \quad \mbox{and} \quad \dot p_t = - \pdd Aq(q_t,p_t),\] which
are Hamilton's equations if \(A\) is the Hamiltonian.

Now, suppose that \(A(q,p)=T(p)\) is only a function of the momenta, then \(\HV
T=T'(p)\,\pdd{}q\), and Hamilton's equations reduce to the pair \(\dot q_t =
T'(p)\) and \(\dot p_t = 0\) of first-order differential equations with
constant coefficients, with the solution that the momentum is constant, \(p_t =
p_0\), and \(q_t = q_0 + T'(p_0)t\) grows linearly in~\(t\).  The case where
\(A(q,p) = S(q)\) is analogous.  If we have a function \(A(q,p) = H(q,p) = T(p)
+ S(q)\), perhaps the Hamiltonian itself, that can be decomposed into the sum
of a kinetic energy and a potential energy then we can easily integrate either
term separately, and we can use a symplectic integrator to approximate the
integral curves of \(\HV H\) itself.

In fact we have established a stronger result, namely we can find the exact
integral curves of a shadow Hamiltonian \(H_\varepsilon\) that differs from
\(H\) by terms of \(\O(\varepsilon)\) in closed form.  A symplectic integrator
thus not only exactly preserves the symplectic structure but also conserves the
value of \(H\) (the energy) up to order \(\varepsilon\) for arbitrarily long
times: unfortunately the integral curves of \(\HV H\) and \(\HV H_\varepsilon\)
usually diverge from each other after a relatively short time despite this.
This happens even it their equations of motion are not chaotic: symplectic
integrators are very good at conserving energy and phase space volume, but they
are not particularly good in finding the correct trajectory through phase
space.

For HMC applications where we only care about exact reversibility, exact
area-preservation, and good energy conservation we see that symmetric
symplectic integrators meet all the requirements, and the divergence of the
shadow integral curves from the true ones is unimportant.

Given the fundamental 2-form \(\ftf = \d q\wedge\d p\) we may evaluate the
Poisson bracket of two arbitrary 0-forms \(A\) and \(B\), namely
\begin{eqnarray*}
  \lefteqn{\PB AB \defn -\ftf(\HV A,\HV B)} \quad &&  \\
  &=& -(\d q\wedge\d p)\left(\pdd Ap\,\pdd{}q - \pdd Aq\,\pdd{}p,
                           \pdd Bp\,\pdd{}q - \pdd Bq\,\pdd{}p\right) \\
  &=& \pdd Ap\,\pdd Bq - \pdd Aq\,\pdd Bp.
\end{eqnarray*}
For the Hamiltonian \(H(q,p) = T(p) + S(q)\) any integrator constructed from
\(e^{\varepsilon\HV S}\) and \(e^{\varepsilon\HV T}\) steps will conserve a
shadow whose BCH expansion may be expressed in terms of the Poisson brackets
\begin{eqnarray*}
  \PB ST &=& -S'T' \\
  \PB S{\PB ST} &=& -S'\pdd{\PB ST}p = S'^2T'' \\
  \PB T{\PB ST} &=& T'\pdd{\PB ST}q = -S''T'^2
\end{eqnarray*}
and so forth.

For example, the leapfrog integrator \(\left[\exp(\half\dt\HV S)\exp(\dt\HV
  T)\exp(\half\dt\HV S)\right]^{t/\dt}\) is the simplest symmetric symplectic
integrator (there is a variant in which \(\HV S\) and \(\HV T\) are
interchanged).  From \eqref{eq:bchsym} we find that it conserves the shadow
Hamiltonian
\begin{eqnarray*}
  \shadow H &=& T + S - \frac{\dt^2}{24}
      \Bigl(\PB S{\PB S T} + 2\PB T{\PB ST}\Bigr) + \O(\dt^4) \\
    &=& H - \frac{\dt^2}{24}(S'^2T'' - 2S''T'^2) + \O(\dt^4).
\end{eqnarray*}

\subsection{Higher Order Integrators} \label{sec:ho-int}

Let us briefly give some simple examples of more complicated integrators.  The
second order minimum norm integrator \cite{omelyan:2001, Takaishi:1999bi,
  forcrand:2006}~is
\begin{eqnarray*}
  \Bigl[\exp\left(\lambda\dt\HV S\right) \exp\left(\half\dt\HV T\right)
    \exp\left((1-2\lambda)\dt\HV S\right) \qquad\qquad && \\
   \times \exp\left(\half\dt\HV T\right) \exp\left(\lambda\dt\HV S\right)
     \Bigr]^{t/\dt} &&
\end{eqnarray*}
with shadow
\begin{eqnarray*}
  \shadow H = T + S + \dt^2
      \biggl(\frac{6\lambda^2 - 6\lambda + 1}{12} \PB S{\PB S T} && \\
        + \frac{1 - 6\lambda}{24} \PB T{\PB ST}&&\biggr) + \O(\dt^4),
\end{eqnarray*}
and it has the free parameter \(\lambda\) as well as the integration step
size~\(\dt\).

It is interesting to note that if as is usual the kinetic energy is quadratic,
\(T(p) = \half p^2\), then the Poisson bracket \(\PB S{\PB ST} = S'^2\) is
independent of the momentum~\(p\), and thus we can find the integral curve of
its Hamiltonian vector field \(\HV{\PB S{\PB ST}} = -2S'S''\pdd{}p\).  The
corresponding integrator step \(e^{\varepsilon\HV{\PB S{\PB ST}}}\) is called a
\emph{force-gradient} integrator step, because it involves second derivatives
of the potential \(S\).

We can use the force-gradient step to define a force-gradient integrator
\begin{eqnarray*}
  \lefteqn{\biggl[\exp(\rational16\dt\HV S) \exp(\half\dt\HV T)} \qquad && \\
    &\times& \exp\left(\rational1{72}\left[48\,\dt\HV S
      - \dt^3\HV{\PB S{\PB ST}}\right]\right) \\
    && \qquad\qquad\qquad\qquad\times
      \exp(\half\dt\HV T) \exp(\rational16\dt\HV S) \biggr]^{t/\dt}
\end{eqnarray*}
with shadow
\begin{eqnarray}
  \lefteqn{\shadow H = T + S} \quad && \nonumber \\
  && - \frac{\dt^4}{155520} \left(\begin{array}{r}
      41 \PB S{\PB S{\PB S{\PB S T}}} \\[1ex]
    + 36 \PB{\PB ST}{\PB S{\PB S T}} \\[1ex]
    + 72 \PB{\PB ST}{\PB T{\PB S T}} \\[1ex]
    + 84 \PB T{\PB S{\PB S{\PB S T}}} \\[1ex]
    + 126 \PB T{\PB T{\PB S{\PB S T}}} \\[1ex]
    + 54 \PB T{\PB T{\PB T{\PB S T}}} 
  \end{array}
  \right) + \O(\dt^6), \label{eq:fg-shadow}
\end{eqnarray}
where we have chosen the integrator parameters to eliminate all terms of
\(\O(\dt^2)\) in the shadow.  The Poisson bracket \(\PB S{\PB S{\PB ST}}=0\) so
the first and fourth Poisson brackets in \eqref{eq:fg-shadow} are also
identically zero, however formula \eqref{eq:fg-shadow} is valid more generally.
Note that the middle step has combined the Hamiltonian vector fields \(\HV S\)
and \(\HV{\PB S{\PB ST}}\) because they commute.

There is no compelling reason to choose the parameters to eliminate the
\(\dt^2\) errors: in general we should introduce some parameters constrained
only by the conditions that the leading order term in the shadow should be the
original Hamiltonian and that the total step size should be~\(\dt\), and then
adjust these parameters to minimize the cost of our integrator for the specific
problem it is being applied to.  On the other hand, we can build integrators
whose leading error is \(\dt^4\) (or \(\dt^{2n}\) for any \(n\) for that
matter), without requiring force-gradient steps.  Nevertheless, integrators
with force-gradient steps may be cheaper than those without: it would be
surprising if the optimal coefficient of the force-gradient term was exactly
zero.

In HMC for lattice field theory \(H\) and \(\shadow H\) are extensive
quantities, that is they are proportional to the lattice volume~\(V\) for
sufficiently large~\(V\), so the leading error is proportional to \(V\dt^{2n}\)
if \(H-\shadow H=\O(\dt^{2n})\).  In order to keep the Monte Carlo acceptance
rate fixed we therefore need to vary \(\dt\propto V^{-1/2n}\), and as the cost
\(Vt/\dt\) of a trajectory of length \(t\) is proportional to the number of
steps and the volume, we may estimate that the cost varies as~\(V^{1+1/2n}\).
Of course there are many other contributions to the cost that have been
ignored, but for large enough \(V\) this suggests that we want to increase
\(n\).

\section{Hamiltonian Mechanics on Lie Groups} \label{sec:ham-lie}

\subsection{Fundamental 2-Form on a Lie Group}

The cotangent bundle \(T^{*}\G\) over any manifold \(\G\) has a natural
symplectic structure.  For the case where \(\G\) is a Lie group a point in
\(T^{*}\G\) may be written as \((g,\form p)\) where \(g\in\G\) and \(\form p\in
T^{*}\G(g)\) is called the \emph{momentum} or \emph{Liouville} form.  As
explained in \appref{sec:lie-groups}, the vectors in tangent space at the
identity \(T\G(\ident)\) correspond to the Lie algebra of left-invariant vector
fields \(\vec e_i\) on \(\G\), and their dual 1-forms \(\form\theta^i\) satisfy
the Maurer--Cartan equations.  The momentum may be written in the
Maurer--Cartan basis as \(\form p = p_i\form \theta^i\), where \(\form p(\vec
e_j) = p_i\form\theta^i(\vec e_j) = p_i \delta^i_j = p_j\).  We shall choose
the fundamental 2-form to be
\begin{equation}
  \ftf\defn -\d\form p = -\d(p_i\form \theta^i), \label{eq:ftf}
\end{equation}
and using the Maurer--Cartan equations it may be written as \[\ftf =
\form\theta^i\wedge dp_i + \half p_i c^i_{jk} \form\theta^j \wedge\form
\theta^k.\]

If \(F\) is a 0-form on the cotangent bundle \(T^{*}\G\) then the corresponding
Hamiltonian vector field \(\HV F = F^i\vec e_i + \bar F_i \pdd{}{p_i}\) in
\(TT^{*}\G\) is defined by \(\d F = i_{\HV F}\ftf\), or \(\d F(\vec y) =
\ftf(\HV F,\vec y)\) for all vector fields \(y = y^i\vec e_i + \bar y_i
\pdd{}{p_i}\).  Expanding this expression gives
\begin{eqnarray*}
  \d F(\vec y) &=& \vec yF = \vec e_i(F)y^i + \pdd{F}{p_i}\bar y_i \\
    &=& \ftf(\HV F,\vec y) = F^i\bar y_i - y^i\bar F_i + p_ic^i_{jk}F^jy^k,
\end{eqnarray*}
so equating the coefficients of \(y^i\) and \(\bar y_i\) we find \(\vec e_i(F)
= - \bar F_i + p_jc^j_{ki}F^k\) and \(\pdd F{p_i} = F^i\).  We thus find that
the vector field \(\HV F\) is
\begin{equation}
  \HV F = \pdd F{p_i}\vec e_i + \left(p_jc^j_{ki}\pdd F{p_k} 
    - \vec e_i(F)\right)\pdd{}{p_i}. 
  \label{eq:vf}
\end{equation}
From this we can evaluate the Poisson bracket of two arbitrary Hamiltonian
vector fields corresponding to 0-forms \(F\) and \(G\),
\begin{eqnarray}
  \PB FG &\defn& -\ftf(\HV F,\HV G) \nonumber \\
  &=& p_ic^i_{jk}\pdd F{p_j}\pdd G{p_k}
    + \pdd F{p_i}\vec e_i(G) - \pdd G{p_i}\vec e_i(F).
  \label{eq:HV-PB}
\end{eqnarray}

\subsection[Hamiltonian Vector Fields for T and S]{Hamiltonian Vector Fields
  for \(T\) and \(S\)}

For HMC we may take the Hamiltonian to be of the form \(H = T + S\) where the
kinetic energy \(T: T^{*}\G\to\R\) is a function only of the momenta which we
may choose to be of the form
\begin{equation}
  T = \half\CK{\form p}{\form p} = \half p^ip_i \label{eq:ke}
\end{equation}
using the Cartan--Killing metric (\secref{sec:cartan-killing}).  Hence \(\pdd
T{p_i} = p^i\), and the potential energy \(S:\G\to\R\) is a function only of
the group parameters.

For the kinetic and potential energy 0-forms the corresponding vector fields
are thus
\begin{equation}
  \HV T = p^i\vec e_i + c^j_{ki}p_jp^k\pdd{}{p_i} = p^i\vec e_i
  \:\:\mbox{and}\:\:\HV S = - \vec e_i(S) \pdd{}{p_i}
  \label{eq:ST-vector-fields}
\end{equation}
using~\eqref{eq:vf}, where we have made use of the total antisymmetry of the
structure constants for a semisimple Lie algebra, \(c^j_{ki}p_jp^k =
c_{jki}p^jp^k = 0\).

\subsection[Poisson Brackets of S and T]{Poisson Brackets of \(S\) and \(T\)}

We may compute the Poisson brackets of \(S\) and \(T\) from~\eqref{eq:HV-PB}
\begin{eqnarray}
  \PB ST &=& -p^i\vec e_i(S) \label{eq:ST} \\
  \PB S{\PB ST} &=& \vec e^i(S)\vec e_i(S) \label{eq:SST} \\
  \PB T{\PB ST} &=& -p^ip^j\vec e_i\vec e_j(S) \nonumber \\
  \PB T{\PB S{\PB ST}} &=& 2p^i\vec e_i\vec e_j(S)\vec e^j(S) \nonumber \\
  \PB S{\PB S{\PB ST}} &=& 0 \nonumber \\
  \PB T{\PB T{\PB ST}} &=& -p^ip^jp^k\vec e_i\vec e_j\vec e_k(S) \nonumber \\
  \PB T{\PB T{\PB S{\PB ST}}}
    &=& 2 p^i p^j \vec e_i\vec e_j\vec e_k(S)\vec e^k(S) \nonumber \\
    &&  + 2 p^i p^j \vec e_i\vec e_k(S)\vec e_j\vec e^k(S) \nonumber \\
  \PB{\PB ST}{\PB T{\PB ST}} &=& c^k_{ij} p^i p^\ell \vec e^j(S)
    \left[\vec e_k\vec e_\ell(S) + \vec e_\ell\vec e_k(S)\right] \nonumber \\
    && + p^i p^j \Bigl[\vec e^k(S)\vec e_k\vec e_i\vec e_j(S) \nonumber \\
    && \qquad\qquad - \vec e^k\vec e_i(S) \vec e_k\vec e_j(S) \nonumber \\
    && \qquad\qquad - \vec e_i\vec e^k(S) \vec e_k\vec e_j(S)\Bigr] \nonumber \\
  \PB T{\PB S{\PB S{\PB ST}}} &=& 0 \nonumber \\
  \PB{\PB ST}{\PB S{\PB ST}}
    &=& -2\vec e^i(S)\vec e^j(S)\vec e_i\vec e_j(S) \nonumber \\
  \PB T{\PB T{\PB T{\PB ST}}}
    &=& -p^ip^jp^kp^\ell\vec e_i\vec e_j\vec e_k\vec e_\ell(S) \nonumber \\
  \PB S{\PB S{\PB S{\PB ST}}} &=& 0 \label{eq:SSSST}
\end{eqnarray}
Observe that according to equation~\eqref{eq:SST} \(\PB S{\PB ST}\) does not
depend on the momentum, so just as in \secref{sec:ho-int} we can use it to
define a force-gradient integrator step corresponding to the Hamiltonian vector
field
\begin{equation}
  \HV{\PB S{\PB ST}}
    = -\vec e_i\Bigl(\vec e^j(S)\vec e_j(S)\Bigr) \pdd{}{p_i}.
  \label{eq:hv-sst}
\end{equation}

\subsection{Representations} \label{sec:representations}

If \(U:\G\to\Gl(n,\C)\defn\Aut\C^N\) is a matrix representation of \(\G\) then
it satisfies \(U(gh) = U(g)U(h)\) for all \(g,h\in\G\).  We may view any matrix
element \(U_{ab}\) of the representation as a complex valued 0-form as it is
well-defined over the entire group manifold.  The left action \(L_g:h\mapsto
gh\) induces the map \({L_g}_{*}: U_{ab}\mapsto U_{ab}\circ L_g\) according to
the definition given in \secref{sec:induced-maps}, so \(({L_g}_{*}U_{ab})(h) =
U_{ab}(gh) = [U(g)U(h)]_{ab} = \sum_{c=1}^n U_{ac}(g)U_{cb}(h)\) for all \(h\),
or equivalently \({L_g}_{*} U_{ab} = \sum_{c=1}^n U_{ac}(g)U_{cb}\).  In other
words the map \({L_g}_{*}\) takes the 0-form \(U_{ab}\) to a linear combination
of 0-forms \(U_{cb}\) with coefficients \(U_{ac}(g)\in\C\).  We can express
this more succinctly by considering \(U\) to be a matrix-valued 0-form, whence
\({L_g}_{*} U = U(g)U\).

Application of the vector field \(\vec e_i\) to \(U\) gives a matrix-valued
0-form \(\vec e_iU\) whose value at some point \(g\in\G\) is \(\vec e_iU(g) =
{L_g}_{*}\vec e_iU(\ident)\).  \(\vec e_i\) is left-invariant \(L_g^{*}\vec e_i
= \vec e_i\), so we have \({L_g}_{*}\vec e_iU = {L_g}_{*} L_g^{*} \vec e_iU =
{L_g}_{*} {L_{g^{-1}}}_{*} \vec e_i {L_g}_{*} U = \vec e_i {L_g}_{*}U = \vec
e_i U(g)U = U(g)\vec e_iU\).  This allows us to evaluate \(\vec e_iU\) at any
point \(g\) in terms of the value of \(\vec e_iU\) at the identity.  Defining
the \emph{generators} of the representation as \(T_i \defn \vec e_iU(\ident)\),
we obtain \(\vec e_iU(g) = U(g)\vec e_iU(\ident) = U(g)T_i\) or more succinctly
\(\vec e_iU = UT_i\).

As on the one hand \(\CT{\vec e_i}{\vec e_j}U = c^k_{ij} \vec e_kU = c^k_{ij}
UT_k\), and on the other \(\CT{\vec e_i}{\vec e_j}U = \vec e_i\vec e_jU - \vec
e_j\vec e_iU = \vec e_iUT_j - \vec e_jUT_i = UT_iT_j - UT_jT_i = U\CT{T_i}
{T_j}\), we see that the generators must satisfy the commutation relations
\(\CT{T_i}{T_j} = c^k_{ij} T_k\) upon multiplying on the left by \(U^{-1}\).

Unfortunately the usual convention \protect{\cite{gottlieb87a, kennedy88b}} is
that the derivative of a link variable is
\begin{equation}
  \vec e_iU = -T_iU, \label{eq:eiu}
\end{equation}
and this is used in most computer implementations.  This arises from
considering right-invariant vector fields.  Briefly, the right action on a
group is defined by \(R_g:h\mapsto hg\), and the induced maps by \({R_g}_{*}U =
U\circ R_g\) and \(R_g^{*}\vec e_i = {R_{g^{-1}}}_{*}\vec e_i {R_g}_{*}\).  If
we assume that \(\vec e_i\) is \emph{right-invariant} then it satisfies
\(R_g^{*}\vec e_i=\vec e_i\), and following an argument completely analogous to
that in the text we find \(\vec e_iU(g) = {R_g}_{*}\vec e_iU(\ident)\) since
\(g = R_g\ident = L_g\ident\) and \({R_g}_{*}\vec e_iU = \vec e_iUU(g)\).  We
then have to define the generators by \(\vec e_iU(\ident) = -T_i\), leading to
\(\vec e_iU = -T_iU\).  We must include the minus sign in the definition of the
generators for right-invariant vector fields satisfying \(\CT{\vec e_i}{\vec
  e_j}=c_{ij}^k\vec e_k\) as otherwise they would not satisfy the commutation
relations \(\CT{T_1}{T_j}=c_{ij}^kT_k\).  In fact, the usual convention
erroneously omits the minus sign, but as the commutation relations are used to
derive the Maurer--Cartan equations, and thus our fundamental 2-form, the sign
is significant when computing high order Poisson brackets.



\subsection{Equations of Motion}

The equations of motion are most naturally expressed in terms of Lie
derivatives~(\secref{sec:lie-derivatives}).  The Lie derivative \(\L_{\vec v}
\vec T\) of a tensor field \(\vec T\) is its derivative along the integral
curves of the vector field \(\vec v\), and the definition of the Lie derivative
given in \eqref{eq:lie-deriv-0-form}, \eqref{eq:lie-deriv-vector},
and~\eqref{eq:lie-deriv-k-form} implicitly provides the differential equations
defining these integral curves.  If \(\vec v=\HV H\) is the Hamiltonian vector
field for the Hamiltonian function then these are just Hamilton's equations,
and we will write \(\dot\vec T \defn \L_{\HV H}\vec T\).

For the case of matrix representations we consider the matrix elements to be
0-forms as we did in \secref{sec:representations} so we may use
equation~\eqref{eq:lie-deriv-0-form} to obtain \(\dot U \defn \L_{\HV H}U = \HV
HU\) and \(\dot P \defn \L_{\HV H}P = \HV HP\) where \(U\) is a matrix
representation of an element of \(\G\) and \(P\defn p^iT_i\) the corresponding
matrix representation of the momentum in the Lie algebra.  Taking \[\HV H = \HV
T+\HV S = p^i\vec e_i - \vec e_i(S) \pdd{}{p_i}\] with the explicit forms from
\eqref{eq:ST-vector-fields}, and using the relation \(\vec e_i(U) = -T_iU\) of
\eqref{eq:eiu}, we find
\begin{eqnarray*}
  \dot U &=& \HV TU = p^i\vec e_i(U) = -p^iT_iU = -PU \\
  \dot P &=& \HV SP = -\vec e_i(S) \pdd P{p_i} = -\vec e_i(S)T^i = -F_1
\end{eqnarray*}
where we have introduced the quantity \(F_1 \defn \vec e_i(S)T^i\) (q.v.,
equation~\eqref{eq:vector-fields}).  The solution of these equations for
separate \(U\) and \(P\) updates (i.e., for a symplectic integrator) are
\[U(t)=\exp(-P t)\,U(0) \qquad\mbox{and} \qquad P(t)=P(0)-tF_1.\]

The equations of motion for the force-gradient Hamiltonian vector field
of~\eqref{eq:hv-sst} is
\begin{eqnarray}
  \dot P &=& \HV{\PB S{\PB ST}}P
    = -\vec e_i\Bigl(\vec e_j(S)\vec e^j(S)\Bigr) \pdd P{p_i} \nonumber \\
    &=& -2\,\vec e_i\vec e_j(S) \vec e^j(S) T^i = -G
    \label{eq:fg}
\end{eqnarray}
with \(G \defn \vec e_i\vec e_j(S) \vec e^i(S) T^j\) (q.v.,
equation~\eqref{eq:vector-fields}), since \(\CT{\vec e_i}{\vec e_j}(S) \vec
e^j(S) = c_{kij}\vec e^k(S)\vec e^j(S) = 0\).
				       
\section[Poisson brackets in SU(N)]{Poisson brackets in \(SU(N)\)}
\label{sec:SUN}


In order to compute the Poisson brackets it is useful to express them in terms
of the following set of matrices that are in the representation of the Lie
algebra
\def\1#1#2#3{#2\ifx#1\relax\relax\else\;\defn\;#1\fi
  &\ifx#1\relax\relax\defn\else=\fi&#3T^i\\}
\begin{equation}
  \begin{array}{rcl}
    \1{}{P}{p_i}
    \1{}{F_1}{\vec e_i(S)}
    \1{\vp PF_1}{F_2}{p^j\vec e_j\vec e_i(S)}
    \1{\vp P^2F_1}{F_3}{p^k\vec e_kp^j\vec e_j\vec e_i(S)}
    \1{\vp P^3F_1}{F_4}{p^\ell\vec e_\ell p^k\vec e_kp^j\vec e_j\vec e_i(S)}
    \1{\vp F_1F_1}{G}{\vec e^j(S)\vec e_j\vec e_i(S)}
  \end{array}
  \label{eq:vector-fields}
\end{equation}
\def\1#1#2#3{#3&=&\Tr(#2T_i)/a\\}
\[
  \begin{array}{rcl}
    \1{}{P}{p_i}
    \1{}{F_1}{\vec e_i(S)}
    \1{\vp PF_1}{F_2}{p^j\vec e_j\vec e_i(S)}
    \1{\vp P^2F_1}{F_3}{p^k\vec e_kp^j\vec e_j\vec e_i(S)}
    \1{\vp P^3F_1}{F_4}{p^\ell\vec e_\ell p^k\vec e_kp^j\vec e_j\vec e_i(S)}
    \1{\vp F_1F_1}{G}{\vec e^j(S)\vec e_j\vec e_i(S)}
  \end{array}
\]
where \(\vp P=p^i\vec e_i\) and \(\vp F_1=\vec e^i(S)\vec e_i\) are vector
fields (linear differential operators) corresponding to the matrices \(P\) and
\(F_1\) respectively.  For a lattice field theory \(P, F_i, G, \ldots\) will
also be lattice vectors, so shall call these quantities \emph{{\plv}s}.

To derive more explicit expressions for the desired Poisson brackets it is
useful to use the following identities that hold for the fundamental
representation of the \(\su(N)\) Lie algebra \footnote{We choose to normalize
the traceless anti-Hermitian generators \(T_i\) of the fundamental
representation by \(\Tr(T_iT_j) = a\,\delta_{ij}\), where \(a\) is an arbitrary
(negative) constant.  For \(\su(3)\) the Hermitian Gell-Mann matrices
\(\lambda_i\) satisfy \(\Tr(\lambda_i\lambda_j) = 2\delta_{ij}\), so our choice
corresponds to \(T_i = \sqrt{-a/2}\;i\lambda_i\).  Moreover, our definition of
the kinetic energy is \(T = \half p_ip^i = \Tr(P^2)/2a\), and as we observed in
the introduction changing this normalization corresponds to a scaling of
molecular dynamics time.  One must be careful to take all these factors into
account when comparing computations using different conventions.}, for arbitary
\(N\times N\) matrices \(X,Y,Z\), and \(T_i\in\su(N)\)
\begin{equation}
  c^i_{jk} \Tr(XT^j) \Tr(YT^k) = a\Tr\left(\CT XYT^i\right);
  \label{eq:id1}
\end{equation}
\begin{equation}
  \Tr(XT_i) \Tr(YT^i) = a\left[\Tr(XY) - \frac1N\Tr X\Tr Y\right];
  \label{eq:id2}
\end{equation}
\begin{equation}
  \Tr\CT XY = \Tr(XY - YX) = 0;
  \label{eq:id5}
\end{equation}
and
\begin{eqnarray}
  \Tr\left(\CT XYZ\right) &=& \Tr(XYZ-YXZ) = \Tr(XYZ - XZY) \nonumber \\
    &=& \Tr\left(X\CT YZ\right) = \Tr\left(\CT YZX\right) \nonumber \\
    &=& \Tr\left(\CT ZXY\right); \label{eq:id6}
\end{eqnarray}
from which it follows that
\begin{equation}
  \Tr\left(\CT XYX\right) = \Tr\left(\CT XXY\right) = 0
  \label{eq:id7}
\end{equation}
and
\begin{equation}
  c_{ijk} \Tr(XT^i) \Tr(YT^j) \Tr(ZT^k) = a^2\Tr\left(\CT XYZ\right).
  \label{eq:id8}
\end{equation}

Using \eqref{eq:id8} and~\eqref{eq:id6} we easily see that
\begin{eqnarray}
  c^k_{ij} p^ip^\ell\vec e^j(S) \vec e_\ell\vec e_k(S)
    &=& \frac1{a^3} c_{ijk} \Tr(PT^i) \Tr(F_1T^j) \Tr(F_2T^k) \nonumber \\
    &=& \frac1a \Tr\Bigl(\CT{F_1}{F_2}P\Bigr),
  \label{eq:new1}
\end{eqnarray}
and as \eqref{eq:id1} leads to
\begin{eqnarray}
  p^\ell\CT{\vec e_k}{\vec e_\ell}(S) = p^\ell c^i_{k\ell} \vec e_i(S)
    &=& \frac1{a^2} c_{k\ell i} \Tr(PT^\ell) \Tr(F_1T^i) \nonumber \\
    &=& \frac1a \Tr\left(\CT{P}{F_1}T_k\right)
  \label{eq:new2}
\end{eqnarray}
we find using \eqref{eq:id8} that
\begin{eqnarray}
  \lefteqn{c^k_{ij} p^ip^\ell\vec e^j(S) \CT{\vec e_k}{\vec e_\ell}(S)}
    \qquad && \nonumber \\
    &=& \frac1{a^3} c_{ijk}\Tr(PT^i) \Tr(F_1T^j)
      \Tr\left(\CT{P}{F_1}T^k\right) \nonumber \\
    &=& \frac1a\Tr\left({\CT{F_1}{P}}^2\right).
  \label{eq:new3}
\end{eqnarray}
Combining equations~\eqref{eq:new1} and~\eqref{eq:new3} we obtain
\begin{eqnarray}
  \lefteqn{c^k_{ij}p^ip^\ell\vec e^j(S) \left\{\vec e_k\vec e_\ell(S)
      + \vec e_\ell\vec e_k(S)\right\}} \qquad && \nonumber \\
    &=& \frac1a \Tr\left(2\CT{F_1}{F_2}P + {\CT{F_1}P}^2\right).
  \label{eq:id14}  
\end{eqnarray}

We may also deduce from \eqref{eq:new2} that
\begin{displaymath}
  p^i\vec e_k\vec e_i(S)
    = \frac1a \Tr\left((F_2 - \CT{F_1}P)T_k\right),
\end{displaymath}
and hence
\begin{equation}
  p^ip^j\vec e^k\vec e_i(S) \vec e_k\vec e_j(S)
    = \frac1a \Tr\left(\Bigl(F_2 - \CT{F_1}P\Bigr)^2\right)
  \label{eq:new4}
\end{equation}
and
\begin{equation}
  p^ip^j\vec e_i\vec e^k(S) \vec e_k\vec e_j(S)
    = \frac1a \Tr\left(F_2^2 - F_2\CT{F_1}P\right).
  \label{eq:new5}
\end{equation}
From the identity
\begin{eqnarray*}
  \vec e_k\vec e_i\vec e_j
    &=& \CT{\vec e_k}{\vec e_i}\vec e_j + \vec e_i\CT{\vec e_k}{\vec e_j}
      + \vec e_i\vec e_j\vec e_k \\
    &=& c^\ell_{ki} \CT{\vec e_\ell}{\vec e_j} + c^\ell_{ki}\vec e_j\vec e_\ell
      + \vec e_ic^\ell_{kj}\vec e_\ell + \vec e_i\vec e_j\vec e_k \\
    &=& c^\ell_{ki} c^m_{\ell j}\vec e_m + c^\ell_{ki}\vec e_j\vec e_\ell
      + c^\ell_{kj}\vec e_i\vec e_\ell + \vec e_i\vec e_j\vec e_k
\end{eqnarray*}
we deduce that
\begin{eqnarray}
  && p^ip^j\vec e^k(S)\vec e_k\vec e_i\vec e_j(S) \nonumber \\
  && \qquad = c^\ell_{ki} \vec e^k(S) p^i c^m_{\ell j} p^j\vec e_m(S)
      + 2c^\ell_{ki}p^ip^j\vec e_j\vec e_\ell(S)\vec e^k(S) \nonumber \\
    && \qquad\qquad\qquad
      + p^ip^j\vec e_i\vec e_j\vec e_k(S) \vec e^k(S) \nonumber \\
  && \qquad = \frac1{a^4} c^\ell_{ki} \Tr(F_1T^k) \Tr(PT^i)
      c_{\ell jm} \Tr(PT^j) \Tr(F_1T^m) \nonumber \\
  && \qquad\qquad\qquad
      + \frac2{a^3} c_{ki\ell} \Tr(PT^i) \Tr(F_2T^\ell) \Tr(F_1T^k) \nonumber \\
  && \qquad\qquad\qquad + \frac1{a^2} \Tr(F_3T_k) \Tr(F_1T^k) \nonumber \\
  &&\qquad = -\frac1a \Tr\left({\CT{F_1}P}^2 + 2\CT{F_1}{F_2}P - F_1F_3\right).
  \label{eq:new6}
\end{eqnarray}

We thus obtain the following expressions for the desired Poisson brackets
\begin{eqnarray*}
  \PB ST &=& 
    -\Tr(F_1P)/a \\
  \PB S{\PB ST} &=& 
    \Tr(F_1^2)/a \\
  \PB T{\PB ST} &=& 
    -\Tr(F_2P)/a \\
  \PB T{\PB S{\PB ST}} &=& 
    2\Tr(F_1F_2)/a \\
  \PB S{\PB S{\PB ST}} &=& 0 \\
  \PB T{\PB T{\PB ST}} &=& 
    -\Tr(F_3P)/a \\
  \PB T{\PB T{\PB S{\PB ST}}}
    &=& 2\left\{ \Tr(F_1F_3) + \Tr(F_2^2)\right\}/a \\
  \PB{\PB ST}{\PB T{\PB ST}} 
    &=& -\Tr\Bigl(3\CT{F_1}{F_2}P+{\CT{F_1}P}^2 \\
    && \qquad\qquad\qquad -F_1F_3+2F_2^2\Bigr)/a \\
    && \mbox{using \eqref{eq:id14}, \eqref{eq:new4}, \eqref{eq:new5},
	and \eqref{eq:new6}} \\
  \PB T{\PB S{\PB S{\PB ST}}} &=& 0 \\
  \PB{\PB ST}{\PB S{\PB ST}} &=& 
    -2\Tr(F_1G_1)/a \\
  \PB T{\PB T{\PB T{\PB ST}}} &=&
    -\Tr(F_4P)/a \\
  \PB S{\PB S{\PB S{\PB ST}}} &=& 0.
\end{eqnarray*}

\section{{\Plv}s and Towers} \label{sec:towers}

\subsection{Single Link Updates}

We now consider how to evaluate the {\plv}s of~\eqref{eq:vector-fields}.  This
is particularly simple to do in the case where there is only a single link
variable \(U\), or on a lattice if we choose to only update a single link by
setting the coefficient of the kinetic energy to zero everywhere else as
described in~\secref{sec:multiple-links}.  In this case the potential is of the
form \footnote{We consider the case where the action is linear in \(U\) without
loss of generality, because if it occurs multiple times we can transform it
into a form linear in its tensor product, which can be reduced into a sum of
irreducible representations.  For example, the action \(S = \Re\Tr(UXUX') =
\Re\Tr[(U\otimes U)X'']\) where \((U\otimes U')_{ij,k\ell} = U_{ik}U_{j\ell}\)
and \(X''_{k\ell,ij} = X_{kj}X'_{\ell i}\) are \(N^2\times N^2\) matrices, and
\(U\otimes U\) can be reduced into as sum of two irreducible representations
acting on vectors of dimensions \(\half N(N-1)\) and \(\half N(N+1)\).}  \(S =
\Re\Tr(UX)\) where \(X\) is some constant \(N\times N\) matrix, which in
general is neither in the group nor its algebra.  On a lattice where we are
only updating a single link \(X\) is constructed out of products of other link
variables, which are themselves constant in molecular dynamics time.  We find
\(F_1 = \vec e_i(S)T^i = \Re\Tr\left(\vec e_i(U)X\right)T^i = -\Re\Tr(T_iUX)T^i
= -\Re\Tr(UXT_i)T^i = -a\TAHP(UX)\) where \(\TAHP\) projects onto the Lie
algebra, i.e., the traceless anti-Hermitian part for~\(\su(N)\).  Likewise, 
\(F_2 = \vp PF_1 = p^j\vec e_j\vec e_i(S)T^i = \Re\Tr\Bigl(p^j\vec e_j\vec
e_i(U)X\Bigr)T^i = \Re\Tr\Bigl(p^j\vec e_j(-T_iU)X\Bigr)T^i = -\Re\Tr\Bigl(T_i
p^j\vec e_j(U)X\Bigr)T^i = \Re\Tr(T_ip^jT_j UX)T^i = \Re\Tr(PUXT_i)T^i = a
\TAHP(PUX)\), and so forth for the remaining quantities 
in~\eqref{eq:vector-fields}
\begin{displaymath}
  \begin{array}{l*{3}{@{\:=\:}r}}
    F_1 & \multicolumn{1}{l}{} & -\Re\Tr(UXT_i)T^i    & -a\TAHP(UX), \\
    F_2 & \vp PF_1             & \Re\Tr(PUXT_i)T^i    & a\TAHP(PUX), \\
    F_3 & \vp P^2F_1           & -\Re\Tr(P^2UXT_i)T^i & -a\TAHP(P^2UX), \\
    F_4 & \vp P^3F_1           & \Re\Tr(P^3UX T_i)T^i & a\TAHP(P^3UX), \\
    G   & \vp F_1F_1           & \Re\Tr(F_1UXT_i)T^i  & a\TAHP(F_1UX).
  \end{array}
\end{displaymath}

\subsection{Lattice Updates}

When we have many links we trivially generalize the definition of the
fundamental 2-form \eqref{eq:ftf} to become sums over all links
\begin{eqnarray*}
  \ftf &=& - \sum_\ell \d\form p(\ell)
    = -\sum_\ell \d\Bigl(p_i(\ell)\form \theta^i(\ell)\Bigr) \\
    &=& \sum_\ell \left(\form\theta^i(\ell)\wedge dp_i(\ell) 
      + \half p_i(\ell) c^i_{jk}
        \form\theta^j(\ell)\wedge\form\theta^k(\ell)\right).
\end{eqnarray*}
We can compress the notation by letting indices such as \(i\) also range over
all links: that is \(i\to(i,\ell_i)\) and the implicit sum over the basis of
the Lie algebra \(\sum_i\) becomes an implicit double sum \(\sum_{\ell_i}
\sum_i\).  Of course, we also need to augment the structure constants
\(c_{ij}^k\to c_{(i,\ell_i)(j,\ell_j)}^{(k,\ell_k)} \defn c_{ij}^k
\delta_{\ell_i}^{\ell_k} \delta_{\ell_j}^{\ell_k}\) since the Maurer--Cartan
equations do not mix links.  Similarly, the kinetic energy \eqref{eq:ke}
becomes
\begin{eqnarray*}
  T &=& \half \sum_\ell c(\ell) \CK{\form p(\ell)}{\form p(\ell)}
    = \half \sum_\ell c(\ell) g_{ij} p^i(\ell)p^j(\ell) \\
    &=& \half \sum_\ell c(\ell) p_i(\ell)p^i(\ell)
\end{eqnarray*}
where, as discussed in \secref{sec:multiple-links}, it is convenient to
introduce a separate coefficient \(c(\ell)\) in the kinetic energy for each
link.  We can extend our compressed notation by implicitly associating a factor
of \(c(\ell)\) with each occurence of the augmented Cartan--Killing metric,
\(g_{ij}\to g_{(i,\ell_i)(j,\ell_j)}\defn c(\ell_i)g_{ij}\delta_{\ell_i\ell_j}\) and
hence with every contracted index~\(i\).  With these conventions the definition
looks like \eqref{eq:ftf} and \eqref{eq:ke} again. The sums propagate to the
Poisson brackets where the implicit sums over the indices in
equations~\eqref{eq:ST}--\eqref{eq:SSSST} also become sums over all links,
although second derivatives such as \(\vec e_i\vec e_j(S)\) have bounded
support for an ultralocal action.  It is important to note that the implicit
factor of \(c_{\ell_i}\) associated with contracted indices means that even
though \(\PB S{\PB ST}\) does not depend on any momentum it still has a factor
of \(c(\ell)\) associated with each term.  If we set \(c(\ell') = \delta_{\ell
  \ell'}\) then only link \(\ell\) will appear in equations \eqref{eq:SST}
and~\eqref{eq:fg}, and the force-gradient integrator will therefore only act on
that link.

\subsection{Towers}

The situation would seem to be much more difficult when we want to update all
of the link variables simultaneously; derivatives like \(\vec e_{i_1}\ldots\vec
e_{i_k}(S)\) depend on \(k\) links and it might appear that it will be
prohibitively expensive to compute them.  Fortunately we can avoid this
combinatorial explosion; the key observation is that all the Poisson brackets
and forces only depend on the {\plv}s, and these have only a single free
lattice index.  To make use of this we introduce towers of {\plv}s: a
\emph{tower} \(T(A,B)\) is a an array of {\plv}s \(T(A,B)_i = \vp A^iB\) where
\(A\) is a {\plv}, \(\vp A\) is the vector field associated with it, \(B\) is a
sum of products of gauge links, and the index \(i\in \{0, \ldots, n-1\}\) where
we call \(n\) the \emph{height} of the tower.

The {\plv}s in \eqref{eq:vector-fields} may be constructed from the two towers
\(T(P,B)\) and \(T(F_1,B)\) of heights four and two, where \(B\) is the stencil
of the action~\(S\).  The \emph{stencil} is the collection of all paths in the
action that start with a given link.  For example, in the case of lattice gauge
theory without dynamical fermions the action is a sum of Wilson loops, each
Wilson loop being the trace of the product of gauge links around a closed loop.
This means we can write the action as \(S = \Re\Tr(U_\ell\staple)+S_0\) where
the \emph{staple} \(\staple\) is the sum of products of gauge links along paths
connecting the end of the link \(\ell\) to its beginning, and \(S_0\) is
independent of \(U_\ell\), as in~\secref{sec:pf-forces}.  The stencil in this
case is \(U_\ell\staple\).  This is familiar from the computation of the force
acting on~\(U_\ell\)
\begin{eqnarray}
  F_1(\ell) &=& \vec e_i(S)T^i
    = \vec e_i\Bigl(\Re\Tr U_\ell\staple\Bigr)T^i \nonumber \\
    &=& \Re\Tr\Bigl(e_i(U_\ell)\staple\Bigr)T^i
    = \Re\Tr(-T_iU_\ell\staple) T^i \nonumber \\
    &=& -\Re\Tr(U_\ell\staple T_i)T^i
    = -a\TAHP(U_\ell\staple). \label{eq:stencil}
\end{eqnarray}
The thing to notice here is that we are computing the force on the gauge link
\(U_\ell\) so the index \(i\) is really the pair \((i,\ell)\), and thus \(\vec
e_i(U_{\ell'}) = 0\) for any other link \(\ell'\neq\ell\): in particular,
\(\vec e_i(\staple) = 0\), \(\vec e_i(S_0) = 0\), and \(\vec e_i(U_\ell\staple)
= \vec e_i(U_\ell)\staple\).  Naturally, we want to compute the force acting on
every gauge link, and so the stencil computation of \eqref{eq:stencil} must be
carried out separately for each link: these computations can be done in
parallel if desired.

In order to compute the {\plv} \(\vp A^jF_1 = \vp A^j\vec e_i(S)T^i\) we
proceed as follows:
\begin{eqnarray*}
  \vp A^jF_1(\ell) &=& \vp A^j\vec e_i(S)T^i
    = \vp A^j \Re\Tr(-T_iU_\ell\staple)T^i \\
    &=& -\Re\Tr\Bigl(T_i\vp A^j(U_\ell\staple)\Bigr) T^i \\
    &=& -\Re\Tr\Bigl(T_iT(A,U_\ell\staple)_j\Bigr) T^i \\
    &=& -a\TAHP(T(A,U_\ell\staple)_j\Bigr).
\end{eqnarray*}
This is easy to do if we can compute the tower \(T(A,U_\ell\staple)\) on the
stencil \(U_\ell\staple\).

\subsection{Algebra of Towers}

It is simple to construct the tower \(T(A,B)\) when \(B\) is a single gauge
link \(U\); we have \(T(A,U)_j = \vp A^jU = (-A)^jU\).  This follows from the
definitions \(T(A,U)_0 = U\) and \(\vp A = a^i\vec e_i\) where \(A = a^iT_i\),
so by induction \(T(A,U)_{j+1} = \vp A^{j+1}U = \vp A(\vp A^jU) = \vp A(-A)^jU
= a^i\vec e_i\Bigl((-A)^jU\Bigr) = (-A)^ja^i\vec e_i(U) = (-A)^ja^i(-T_iU) =
(-A)^{j+1}U\).  Indeed, this corresponds to a convenient recursive way of
constructing the tower, \(T(A,U)_{j+1} = (-A)T(A,U)_j\).

If \(B\) is the product \footnote{Here we use the symbol \(\cdot\) to emphasise
multiplication operations.  Elsewhere we use juxtaposition to indicate
multiplication.} of two stencils \(B_1\cdot B_2\) then we may use the Leibniz
rule for the derivation \(\vp A\), \(\vp A(B_1\cdot B_2) = \vp AB_1 \cdot B_2 +
B_1\cdot\vp AB_2\), or more generally \[\vp A^j(B_1\cdot B_2) = \sum_{k=0}^j
{j\choose k} \vp A^kB_1\cdot\vp A^{j-k}B_2.\] The tower on the product
\(B_1\cdot B_2\) is thus the product of the tower on \(B_1\) with that on
\(B_2\), \(T(A,B_1\cdot B_2) = T(A,B_1) \cdot T(A,B_2)\), where the product is
defined by \footnote{The symbol \(\cdot\) on the left denotes multiplication of
towers, whereas on the right it denotes matrix multiplication.}  \[\Bigl(
T(A,B_1) \cdot T(A,B_2)\Bigr)_j = \sum_{k=0}^j {j\choose k} T(A,B_1)_k \cdot
T(A,B_2)_{j-k}.\]

The tower on the sum of two stencils \(B_1+B_2\) is even simpler, since \(\vp
A(B_1+B_2) = \vp AB_1 + \vp AB_2\).  We just have \(T(A,B_1+B_2) = T(A,B_1) +
T(A,B_2)\) where \(\Bigl(T(A,B_1) + T(A,B_2)\Bigr)_j = T(A,B_1)_j+T(A,B_2)_j\).

\subsection{Pseudofermion Towers}

The principal advantage of updating all links simultaneously is when we include
the effects of (pseudo)fermions in the dynamics.  As described in
\secref{sec:pf-forces} this entails solving a large linear system to obtain the
quantity \(X = \Mop^{-1}\phi\) needed compute the force (\(\Mop\) being a
lattice Dirac operator) and it is worthwhile to reuse this solution to update
many links.

We therefore need to compute towers for stencils that include outer products
such as \(X\otimes X^\dagger\).  This may be done by computing the tower
\(T(A,X)\) on \(X = \Mop^{-1}\phi\).  Observe that \(\vp A\phi=0\) as the
pseudofermion lattice (site) vector \(\phi\) does not depend on \(U\) --- we
want to follow the molecular dynamics evolution of the gauge links and momenta
in the presence of a fixed pseudofermion background.  Using the Leibniz rule we
get \(0 = \vp A(\phi) = \vp A(\Mop \Mop^{-1}\phi) = \vp A(\Mop)\Mop^{-1}\phi +
\Mop \vp A(\Mop^{-1}\phi)\) so \(\vp A(\Mop^{-1}\phi) = -\Mop^{-1}\vp A(\Mop)
\Mop^{-1}\phi\).  To use this for a tower of arbitrary height we generalize
this to
\begin{eqnarray*}
  0 &=& \vp A^j(\Mop\Mop^{-1}\phi)
  = \sum_{k=0}^j {j\choose k} \vp A^{j-k}(\Mop)\vp A^k(\Mop^{-1}\phi) \\
  &=& \Mop\vp A^j(\Mop^{-1}\phi)
    + \sum_{k=0}^{j-1} {j\choose k} \vp A^{j-k}(\Mop)\vp A^k(\Mop^{-1}\phi)
\end{eqnarray*}
for \(j>0\), and thus
\[
  \vp A^j(\Mop^{-1}\phi) = -\Mop^{-1} \sum_{k=0}^{j-1} {j\choose k}
    \vp A^{j-k}(\Mop)\vp A^k(\Mop^{-1}\phi).
\]
This translates into the following recursive definition for the tower on \(X\)
\begin{eqnarray*}
  T(A,X)_0 &=& \Mop^{-1}\phi \\
  T(A,X)_j
    &=& -\Mop^{-1} \sum_{k=0}^{j-1} {j\choose k} T(A,\Mop)_{j-k} T(A,X)_k
\end{eqnarray*}
in terms of the tower \(T(A,\Mop)\) which we already know how to compute.  Note
that we require exactly \(n\) inverses to construct such a tower of height of
height~\(n\).

Yin \cite{Yin:2011sz} has suggested an ingeneous way of performing a
force-gradient update by computing the force twice.  We should not be surprised
that the force-gradient update \(e^{\dt^3\HV{\PB S{\PB ST}}}\) can be computed
out of \(e^{\dt\HV S}\) and \(e^{\dt\HV T}\) steps: recall that according to
the BCH formula the commutator \(C(e^A,e^B) = e^{-A} e^{-B} e^A e^B = e^{[A,B]
  + \cdots}\), hence
\begin{eqnarray*}
  \lefteqn{C\left(e^{\dt\HV S}, C(e^{\dt\HV S}, e^{\dt\HV T})\right)} \quad && \\
    &=& e^{-\dt\HV S} e^{-\dt\HV T} e^{-\dt\HV S} e^{\dt\HV T} e^{\dt\HV S} 
      e^{-\dt\HV T} e^{\dt\HV S} e^{\dt\HV T} \\
    &=& C\left(e^{\dt\HV S}, e^{\dt^2\CT{\HV S}{\HV T} + \O(\dt^3)}\right) 
    = e^{\dt^3\CT{\HV S}{\CT{\HV S}{\HV T}} + \O(\dt^4)} \\
    &=& e^{\dt^3\HV{\PB S{\PB ST}} + \O(\dt^4)}.
\end{eqnarray*}
It is interesting that this can be reduced to only requiring two inverses in
the case where \(T\) is quadratic.  There does not seem to be a way of using
this trick to evaluate Poisson brackets, however.


\section{Conclusions}

We have given a formalism for computing integrators and the corresponding
shadow Hamiltonians for lattice gauge theories, and we have presented explicit
formul\ae\ for the Poisson brackets up to fourth order and for the
force-gradient update step.  We have shown how to express these quantities in
terms of {\plv}s taking their values in the representation of the Lie algebra,
as is needed for the usual formulation of lattice gauge theories, and explained
how these may be computed using towers.  The implementation of towers is
straightforward, as it just requires the substitution of the algebra of towers
for that of the matrices already used in computing the force term.  The
stencils for any action are unchanged, and the method is readily applied to
pseudofermions, smeared actions, and so forth.  The rules for addition,
multiplication, and ``inversion'' of towers are given in a recursive form that
is easy to implement (although a recursive implementation is not necessary).

\section*{Acknowledgements}

We would like to thank B\'alint Jo\'o for implementing towers in Chroma.

Paulo Silva acknowledges support from FCT via grant SFRH/BPD/40998/2007, and
project PTDC/FIS/100968/2008, developed under the initiative QREN financed by
the UE/FEDER through the Programme COMPETE --- ``Programa Operacional Factores
de Competitividade''.

\appendix
\section{Differential Forms} \label{sec:diff-forms}

\subsection{Differential Forms and Wedge Products}

For convenience we give the definition of a few basic operations on
differential forms.  In some local basis \(q:\M\supseteq\U\to\R^n\) a
\(k\)-form \(\form\Omega\in\Lambda^k\) has components \footnote{Our convention
  is that each independent component occurs once in the sum: another convention
  is that each such component occurs \(k!\) times --- once for each permutation
  of its indices.}
\begin{eqnarray*}
  \form\Omega &=& \sum_{1\leq I_1<\cdots< I_k\leq k} \Omega_{I_1\ldots I_k}
    \dq^{I_1} \wedge \cdots \wedge \dq^{I_k} \\
   &=& \frac1{k!} \sum_{i_1,\ldots,i_k=1}^N \Omega_{i_1\ldots i_k}
    \dq^{i_1} \wedge \cdots \wedge \dq^{i_k} \\
   &\defn& \frac1{k!} \Omega_{i_1\ldots i_k}
     \dq^{i_1} \wedge \cdots \wedge \dq^{i_k} \\
   &=& \frac1{k!} \sum_{\pi\in\Sym_k} \Omega_{\pi_1\ldots\pi_k}
    \dq^{\pi_1} \wedge \cdots \wedge \dq^{\pi_k} \\
   &=& \left\langle\Omega_{\pi_1\ldots\pi_k}
     \dq^{\pi_1} \wedge \cdots \wedge \dq^{\pi_k} \right\rangle_{\pi\in\Sym_k}
\end{eqnarray*}
where \(\Sym_k\) is the symmetric group acting on \(1,\ldots,k\), and
\(\langle\cdots\rangle_{\Sym_k}\) indicates the average over elements of the
symmetric group.  The \emph{wedge product} satisfies
\[\begin{array}{rcl@{\qquad\qquad}l}
  \form\alpha\wedge\form\beta &=& (-1)^{kk'}\form\beta\wedge\form\alpha
    & \form\alpha\in\Lambda^k, \form\beta\in\Lambda^{k'} \\
    &&& \mbox{Antisymmetry}; \\
  \form\alpha\wedge\form\beta\wedge\form\gamma
    &=& \multicolumn{2}{l}{\form\alpha\wedge(\form\beta\wedge\form\gamma)
    = (\form\alpha\wedge\form\beta)\wedge\form\gamma} \\
    &&& \mbox{Associativity}.
\end{array}\]
In terms of the components in local coordinates this means that \footnote{For
the other convention the numerical coefficient in this formula is
\(1/(k!k'!)\): \emph{caveat emptor}.}
\begin{eqnarray*}
  \lefteqn{\form\alpha\wedge\form\beta
    = \left\langle\alpha_{\pi_1\ldots\pi_k}\beta_{\pi_{k+1}\ldots\pi_{k+k'}}
      \dq^{\pi_1} \!\!\wedge \cdots \wedge
      \dq^{\pi_{k+k'}}\!\right\rangle_{\pi\in\Sym_{k+k'}}} && \\
    &=& \frac1{(k+k')!} \! \sum_{\pi\in\Sym_{k+k'}} \!\!\!\!
      \alpha_{\pi_1\ldots\pi_k} \beta_{\pi_{k+1}\ldots\pi_{k+k'}} \dq^{\pi_1}
        \!\!\wedge \cdots \wedge \dq^{\pi_{k+k'}}.
\end{eqnarray*}

\subsection{Exterior Derivatives}

The \emph{exterior derivative} \(\d:\Lambda^k\to\Lambda^{k+1}\) is a linear
antiderivation, so
\[\begin{array}{rcl@{\quad}l}
  \d(\form\alpha+\form\beta) &=& \d\form\alpha + \d\form\beta
    & \mbox{Linearity}; \\
  \d(\form\alpha\wedge\form\beta) 
    &=& (\d\form\alpha) \wedge \form\beta
      + (-1)^k \form\alpha\wedge \d\form\beta
    & \form\alpha\in\Lambda^k \\ &&& \mbox{Anti-Leibniz}; \\
  \d^2\form\alpha &=& 0 \\
  \d F(\vec x) &=& \vec xF & F\in\Lambda^0.
\end{array}\]
The exterior derivative \(\d F\) for a 0-form \(F\) is defined to be \(\d
F(\vec x) \defn\vec xF\) for any vector field \(\vec x\): if we evaluate this
in a local coordinate system we find that
\begin{eqnarray*}
  \d F(\vec x) &=& \vec xF = \left(x^i\pdd{}{q^i} \right)F
    = \left(\pdd F{q^i}\right)x^i \\ &=& \left(\pdd F{q^i}\dq^i\right)
      \left(x^j\pdd{}{q^j}\right) = \left(\pdd F{q^i}\dq^i\right)(\vec x),
\end{eqnarray*}
so \[\d F = \pdd F{q^i} \dq^i.\] Likewise, in a local coordinate system the
exterior derivative of a \(k\)-form \(\form\Omega\in\Lambda^k\) is
\begin{eqnarray*}
  \d\Omega &=& \d\left(\frac1{k!} \Omega_{i_1\ldots i_k}
    \dq^{i_1} \wedge \cdots\wedge \dq^{i_k} \right) \\
  &=& \frac1{k!} \pdd{\Omega_{i_1 \ldots i_k}}{q^j}\,
    \dq^j \wedge \dq^{i_1}\wedge\cdots \wedge \dq^{i_k}.
\end{eqnarray*}
This follows from the anti-Leibniz rule \(\d(\alpha\form\beta) = \d\alpha\wedge
\form\beta + \alpha\d\form\beta\) applied to the case where \(\alpha =
\Omega_{i_1\ldots i_k}\in\Lambda^0\) and \(\form\beta=\d q^{i_1}
\wedge\cdots\wedge\d q^{i_k}\) because the second term vanishes (by induction
on \(k\)) using the condition \(\d^2=0\) for the basis forms which are exterior
derivatives of the coordinates \(q^i\), \(\d^2q^i=0\) .

In particular, for a 1-form \(\form\theta\in\Lambda^1\) we have \[\d \form
\theta = \pdd{\theta_i}{q^j} \dq^j\wedge \dq^i,\] so applying the 2-form \(\d
\form \theta\) to two arbitrary vector fields \(\vec x\) and \(\vec y\) gives
\begin{eqnarray}
  \lefteqn{\d\form\theta(\vec x,\vec y) = \pdd{\theta_i}{q^j}(x^jy^i - x^iy^j)}
    \quad && \nonumber \\
  &=& x^j \pdd{}{q^j}\left(\theta_iy^i\right) - x^j\theta_i\pdd{y^i}{q^j}
    - y^j \pdd{}{q^j}\left(\theta_ix^i\right) + y^j\theta_i\pdd{x^i}{q^j} 
    \nonumber \\
  &=& \vec x\form\theta(\vec y) - \vec y\form\theta(\vec x)
    - \theta_i\left[\vec x(y^i) - \vec y(x^i)\right] \nonumber \\
  &=& \vec x\form\theta(\vec y) - \vec y\form\theta(\vec x)
    - \form\theta(\CT {\vec x}{\vec y}).
  \label{eq:d-1-form}
\end{eqnarray}
This provides an elegant coordinate-independent definition of
\(\d\form\theta\) in terms of the \emph{commutator} of the vector fields
\begin{eqnarray}
  \CT{\vec x}{\vec y} &\defn& \vec x\vec y - \vec y\vec x
    = x^i\pdd{}{q^i}y^j\pdd{}{q^j} - y^i\pdd{} {q^i} x^j\pdd{}{q^j}
      \label{eq:commutator} \\
    &=& \left(x^i\pdd{y^j}{q^i} - y^i\pdd{x^j}{q^i}\right) \pdd{}{q^j}
      + (x^iy^j - x^jy^i) \pdd{}{q^i}\pdd{}{q^j}, \nonumber
\end{eqnarray}
which is itself a vector field since the last term involving
second derivatives vanishes by symmetry.  Note that if \(\form\theta\) is
exact, that is \(\form\theta=\d F\), then the identity \(\d^2F(\vec x,\vec y) =
\vec x\d F(\vec y) - \vec y\d F(\vec x) - \d F(\CT {\vec x}{\vec y}) = \vec
x\vec yF - \vec y\vec xF - \CT{\vec x}{\vec y}F = 0\) holds automatically.

For an arbitrary \((k-1)\)-form \(\form\Omega\in\Lambda^{k-1}\) we may derive
the corresponding identity,
\begin{eqnarray*}
  \lefteqn{\d\form\Omega(\vec x_1,\ldots,\vec x_k) =
    \sum_{i=1}^k (-1)^{i+1} \vec x_i\form\Omega(\vec x_1,\ldots,
      \hat{\vec x_i},\ldots, \vec x_k)} && \\ 
    && -\!\!\! \sum_{1\leq i<j\leq k} \!\!\!\!
      (-1)^{i+j+1}\form\Omega(\CT{\vec x_i}{\vec x_j},
      \vec x_1, \ldots,\hat{\vec x_i},\ldots,\hat{\vec x_j},\ldots,\vec x_k)
\end{eqnarray*}
where \(\hat\vec x\) indicate that the variable \(\vec x\) is omitted.
We observe that for \(k=3\) the invariant expression for the exterior
derivative is
\begin{eqnarray}
  \lefteqn{\d\ftf(\vec x,\vec y,\vec z)
  = \vec x\ftf(\vec y,\vec z) - \vec y\ftf(\vec x,\vec z)
    + \vec z\ftf(\vec x,\vec y)} \qquad && \nonumber \\
  && - \ftf(\CT{\vec x}{\vec y},\vec z) + \ftf(\CT{\vec x}{\vec z},\vec y)
    - \ftf(\CT{\vec y}{\vec z},\vec x).
  \label{eq:d-2-form}
\end{eqnarray}

\subsection{Interior Products}

The \emph{interior product} \(i:T\M\times\Lambda^k\to\Lambda^{k-1}\) is the
operation that inserts a vector as the first argument of a \(k\)-form to yield
a \(k-1\)-form.  It is formally defined by the axioms
\[
  \begin{array}{rcl@{\quad}rr}
    i_{\vec x}(\form\alpha + \form\beta) 
      &=& i_{\vec x}\form\alpha + i_{\vec x}\form\beta &
      \form\alpha,\form\beta\in\Lambda^k & \mbox{Linearity}; \\
    i_{\vec x}(\form\alpha\wedge\form\beta)
      &=& \multicolumn{3}{l}{i_{\vec x}(\form\alpha)\wedge\form\beta
        + (-1)^k\form\alpha\wedge i_{\vec x}\form\beta} \\
      \multicolumn{4}{r}{\form\alpha\in\Lambda^k,\form\beta\in\Lambda^{k'} }
        & \mbox{Anti-Leibniz}; \\
    i_{\vec x}F &=& 0 && F\in\Lambda^0 \\
      \multicolumn{4}{l}{i_{\vec x}\Omega(\vec x_1,\ldots,\vec x_{k-1})
      = \Omega(\vec x, \vec x_1, \ldots, \vec x_{k-1})} & \Omega\in\Lambda^k; \\
    i_{\vec x}^2 &=& 0
\end{array}
\]
so we see that it too is a linear antiderivation.

\subsection{Induced Maps} \label{sec:induced-maps}

If \(\sigma:\M\to\M'\) is a diffeomorphism, then there is a natural induced map
\(\sigma_{*}:\Lambda^0(\M')\to\Lambda^0(\M)\) defined by \(\sigma_{*}f:
p\mapsto f(\sigma p)\) for all \(f\in\Lambda^0(\M')\) and \(p\in\M\).  This map
may also be written as \(\sigma_{*}f = f\circ\sigma\), and is called a
\emph{pull-back}.  Another way of saying this is that the following diagram
commutes
\begin{displaymath}
  \begin{array}{ccc}
    \M & \stackrel{\sigma}{\longrightarrow} & \M' \\
    \scriptstyle{\sigma_{*}f}\searrow && \swarrow\scriptstyle{f} \\
    & \R &
  \end{array}
\end{displaymath}

If \(\sigma^{-1}\) exists then there is a corresponding pull-back map
\((\sigma^{-1})_{*}\), and it satisfies the relation \((\sigma^{-1})_{*}
\sigma_{*}f = (\sigma^{-1})_{*}(f\circ\sigma) = f\circ\sigma\circ\sigma^{-1} =
f\), and thus we see that \((\sigma^{-1})_{*} = (\sigma_{*})^{-1}\), and we
may denote both of these unambiguously as \(\sigma^{-1}_{*}\).

If \(\vec x\in T\M\) is a vector field on \(\M\) then there may be a
\emph{push-through} map \(\sigma^{*}:T\M\to T\M'\) defined by \(\sigma^{*}\vec
x = \sigma^{-1}_{*}\circ\vec x\circ\sigma_{*}\) if this exists.  For any \(f\in
\Lambda^0(\M')\) and \(p\in\M\) this means that \(\left.  \sigma^{*}\vec x(f)
\right|_{\sigma p} = \left.\vec x(\sigma_{*}f) \right|_p\).  The corresponding
commutative diagram is
\begin{displaymath}
  \begin{array}{ccc}
    \Lambda^{0}(\M) & \stackrel{\sigma_{*}}{\longleftarrow} 
      & \Lambda^{0}(\M') \\
    \vcenter{\llap{$\scriptstyle{\vec x}$}} \Big\downarrow 
      && \Big\downarrow\vcenter{\rlap{$\scriptstyle{\sigma^{*}\vec x}$}} \\
    \Lambda^{0}(\M) & \stackrel{\sigma^{-1}_{*}}{\longrightarrow} &
      \Lambda^{0}(\M').
  \end{array}
\end{displaymath}
The existence of the diffeomorphism \(\sigma^{-1}:\M'\to\M\) is a sufficient
but not necessary condition for \(\sigma^{-1}_{*}\) and hence \(\sigma^{*}\) to
be well-defined.

We may define further induced maps \footnote{One must be careful with the
notation introduced here, as there are a whole family of mappings that we have
given the same name, \(\sigma_{*}:\Lambda^{k}(\M') \to\Lambda^{k}(\M)
\quad\forall k\), and the equation \(\sigma_{*}\form\theta = \sigma_{*}\circ
\form\theta \circ\sigma^{*}\) involves two of them.  If we were to call these
induced mappings on forms \(\sigma^k_{*}:\Lambda^{k}(\M')\to \Lambda^{k}(\M)\)
then the equation is less ambiguous, \(\sigma^1_{*} \form \theta = \sigma^0_{*}
\circ\form\theta \circ\sigma^{*}\).}  such as the pull-back of one-form fields
\(\sigma_{*}:\Lambda^1(\M') \to \Lambda^1(\M)\) as \(\sigma_{*}
\form\theta=\sigma_{*}\circ\form\theta\circ\sigma^{*}\),
\begin{displaymath}
  \begin{array}{ccc}
    T\M & \stackrel{\sigma^{*}}{\longrightarrow} & T\M' \\
    \vcenter{\llap{$\scriptstyle{\sigma_{*}\form\theta}$}} \Big\downarrow 
      && \Big\downarrow\vcenter{\rlap{$\scriptstyle{\form\theta}$}} \\
    \Lambda^{0}(\M) & \stackrel{\sigma_{*}}{\longleftarrow} & \Lambda^{0}(\M'),
  \end{array}
\end{displaymath}
and so forth.

In the special case where \(\sigma:\M\to\M\) is an autodiffeomorphism then the
push-through maps always exist.

\subsection{Lie Derivatives} \label{sec:lie-derivatives}

Suppose now that we have a smooth one-parameter family of diffeomorphisms
\(\sigma:\R\times\M\to\M\), which we will also write as \(\sigma_t:\M\to\M\).
Using this map we can define a derivative with respect to the parameter \(t\),
which is called a \emph{Lie derivative}.  For any 0-form \(F\) we define
\begin{equation}
  \L_{\vec v}F \defn \left.\dd{({\sigma_t}_{*}F)}t\right|_{t=0} 
    = \left.\dd{(F\circ\sigma_t)}t\right|_{t=0} = \vec vF
  \label{eq:lie-deriv-0-form}
\end{equation}
where \(\vec v\) is the linear differential operator --- the vector field ---
that is tangential to the curves \(\sigma(t,p)\) passing through
\(\sigma(0,p)=p\in \M\) at \(t=0\).

The Lie derivative of a vector field \(\vec y\in T\M\) can be deduced from the
requirement that \(\L_{\vec x}\) be a derivation \[\L_{\vec x}(\vec A\otimes\vec
B) = (\L_{\vec x}\vec A)\otimes\vec B + \vec A\otimes\L_{\vec x}\vec B\] for any
tensors \(\vec A\) and \(\vec B\), and that it commutes with contractions
\begin{eqnarray*}
  \L_{\vec x}(\vec yF) &=& (\L_{\vec x}\vec y)F + \vec y(\L_{\vec x}F), \\
  \L_{\vec x}(\form\theta(\vec y))
    &=& (\L_{\vec x}\form\theta)(\vec y) + \form\theta(\L_{\vec x}\vec y)
\end{eqnarray*}
and so forth.  Applying these rules to the 0-form \(\vec yF\) obtained by
applying the vector field \(\vec y\in T\M\) to \(F\in \Lambda^0(\M)\) we have
\(\L_{\vec x}(\vec yF) = \vec x\vec yF\) and also \(\L_{\vec x}(\vec yF) =
(\L_{\vec x}\vec y)F + \vec y(\L_{\vec x}F)\), hence \[(\L_{\vec x}\vec y)F =
\vec x\vec yF - \vec y\vec xF = \CT{\vec x}{\vec y}F\] and, as this holds for
all \(F\),
\begin{equation}
  \L_{\vec x}\vec y = \CT{\vec x}{\vec y}.
  \label{eq:lie-deriv-vector}
\end{equation}

We may apply a similar argument to evaluate the Lie derivative of a 1-form
\(\form\theta\in\Lambda^1(\M)\).  On the one hand \(\L_{\vec x}(\form
\theta(\vec y)) =\vec x\form\theta(\vec y)\), while on the other \(\L_{\vec
x}(\form\theta(\vec y)) = (\L_{\vec x}\form\theta)(\vec y) + \form
\theta(\L_{\vec x}\vec y)\), so using~\eqref{eq:d-1-form}
\begin{eqnarray*}
  \lefteqn{(\L_{\vec x}\form\theta)(\vec y)
  = \vec x\form\theta(\vec y) - \form\theta(\CT{\vec x}{\vec y})
  = \d\form\theta(\vec x,\vec y) + \vec y\form\theta(\vec x)} \quad && \\
  &=& (i_{\vec x}\d\form\theta)(\vec y) + \d(\form\theta(\vec x))(\vec y)
  = (i_{\vec x}\d\form\theta)(\vec y) + (\d i_{\vec x}\form\theta)(\vec y) \\
  &=& (i_{\vec x}\d + \d i_{\vec x})\form\theta(\vec y),
\end{eqnarray*}
hence \[\L_{\vec x}\form\theta = (i_{\vec x}\d + \d i_{\vec x})\form\theta.\]
This suggests that the Lie derivative of any \(k\)-form may be expressed as
\begin{equation}
  \L_{\vec x} = i_{\vec x}\d + \d i_{\vec x}, \label{eq:lie-deriv-k-form}
\end{equation}
and this is indeed the case as the operator \(i_{\vec x}\d + \d i_{\vec x}\) is
a derivation
\begin{eqnarray*}
  \lefteqn{(i_{\vec x}\d + \d i_{\vec x})(\form\alpha\wedge\form\beta)}
    \qquad && \\
  &=& i_{\vec x}\left[(\d\form\alpha)\wedge\form\beta
      +  (-1)^k\form\alpha\wedge\d\form\beta\right] \\
  && \qquad + \d\left[(i_{\vec x}\form\alpha)\wedge\form\beta
      + (-1)^k\form\alpha\wedge i_{\vec x}\form\beta\right] \\
  &=& (i_{\vec x}\d\form\alpha)\wedge\form\beta
    + (-1)^{k+1}(\d\form\alpha)\wedge i_{\vec x}\form\beta \\
  && \qquad + (-1)^k(i_{\vec x}\form\alpha)\wedge\d\form\beta
    + (-1)^{2k}\form\alpha\wedge i_{\vec x}\d\form\beta \\
  && \qquad + (\d i_{\vec x}\form\alpha)\wedge\form\beta
    + (-1)^{k-1} (i_{\vec x}\form\alpha)\wedge\d\form\beta \\
  && \qquad + (-1)^k(\d\form\alpha)\wedge i_{\vec x}\form\beta
    + (-1)^{2k} \form\alpha\wedge\d i_{\vec x}\form\beta \\
  &=& \left[(i_{\vec x}\d + \d i_{\vec x})\form\alpha\right]\wedge\form\beta
    + \form\alpha\wedge(i_{\vec x}\d + \d i_{\vec x})\form\beta
\end{eqnarray*}
for all \(\form\alpha\in\Lambda^k\) and \(\form\beta\in\Lambda^{k'}\), and for
0- and 1-forms \(F\) and \(\form\theta\)
\begin{eqnarray*}
  \L_{\vec x}F &=& \vec xF = \d F(\vec x) = i_{\vec x}\d F + \d i_{\vec x}F, \\
  \L_{\vec x}\form\theta &=& (i_{\vec x}\d + \d i_{\vec x})\form\theta.
\end{eqnarray*}
The second term in the first equation is zero because \(i_{\vec x}F = 0\) by
definition.

\section{Lie Groups} \label{sec:lie-groups}

\subsection{Left-Invariant Forms}

A \emph{Lie group} is a manifold that has a group structure defined by
\(C^\infty\) multiplication \((g,h) \mapsto gh\) and inverse \(g\mapsto
g^{-1}\) operations that satisfy the group axioms
\[\begin{array}{ccccc@{\:\;}l@{\:\;}l}
  g(g'g'') &=& (gg')g'' &\defn& gg'g'' & \forall g,g',g''\in\G 
    & \mbox{Associative} \\
  g^{-1}g &=& gg^{-1} &=& \ident & \forall g\in\G & \mbox{Inverse}
\end{array}\]
with \(\ident\) being the identity element of the group.  If we consider a
point \(g\in\G\) as being ``fixed'' then left multiplication by \(g\) is an
autodiffeomorphism of \(\G\), \(L_g: g'\mapsto gg'\), with \(L_{gh} = L_g\circ
L_h\) by associativity, \(L_g\circ L_hg' = g(hg') = (gh)g' = L_{gh}g'\) for
all \(g'\in\G\).  Clearly \(L_{g^{-1}} = (L_g)^{-1}\) too.

As for any such diffeomorphisms we can define the corresponding pull-back maps
on forms and vectors, \({L_g}_{*}F\defn F\circ L_g\), \(L_g^{*}\vec v\defn
{L_{g^{-1}}}_{*}\circ \vec v\circ{L_g}_{*}\), and \({L_g}_{*}\form\theta\defn
{L_g}_{*}\circ\form\theta\circ L_g^{*}\).  We may use these maps to define
\emph{left-invariant} vector fields and forms; for example, a left-invariant
1-form satisfies the condition \(\form\theta = {L_g}_{*}\form\theta\).

\subsection{Lie Algebra}

The only left-invariant 0-forms are constants, as if \(F = {L_g}_{*}F \quad
(\forall g\in\G)\) then \(F(g) = F(L_g\ident) = {L_g}_{*}F(\ident)=F(\ident)\).

If \(\vec u = L_g^{*}\vec u\) and \(\vec v = L_g^{*}\vec v\) are left-invariant
vector fields in the tangent bundle \(T\G\) then their commutator is also a
vector field, and furthermore it is also left-invariant since \footnote{Note
that \({L_{g^{-1}}}_{*}=(L_g)_{*}^{-1}\)} \(\CT{\vec u}{\vec v}=\CT{L_g^{*}\vec u}
{L_g^{*}\vec v} = \CT{{L_{g^{-1}}}_{*}\circ\vec u \circ{L_g}_{*}}{{L_{g^{-1}}}_{*}
\circ\vec v\circ{L_g}_{*}}={L_{g^{-1}}}_{*}\circ \CT{\vec u}{\vec v}\circ{L_g}_{*}
= L_g^{*}\CT{\vec u}{\vec v}\).  If a left-invariant vector field \(\vec v\)
vanishes at the identity, \(\vec v(F)|_\ident = 0\;(\forall F\in\Lambda^0{\G})\),
then it must be identically zero everywhere, as \(\vec v(F)|_g = \left[\vec
v(F)\circ L_g\right]_\ident = \left[{L_g}_{*} \vec v(F)\right]_\ident =
\left[{L_g}_{*}\circ L_g^{*}\vec v(F)\right]_\ident = \left[{L_g}_{*}
\circ{L_g^{-1}}_{*} \circ\vec v({L_g}_{*} F)\right]_\ident = \left[\vec
v(F\circ L_g)\right]_\ident = 0\).

Consider a set of left-invariant vector fields \(\{\vec e_i\}\) in \(T\G\)
called \emph{generators} whose values at the origin are linearly independent;
any linear combination of the generators with left-invariant (constant)
coefficients is also left-invariant.  Conversely any left-invariant vector
field \(\vec u\) must be a linear combination of this type, since its value at
the origin is \(\vec u|_\ident = \sum_i u^i\vec e_i|_\ident\) with \(u^i\in\R\),
and hence \(\vec u - \sum_i u^i\vec e_i = 0\) everywhere.  Left-invariant
vector fields therefore form a linear space with constant coefficients.  In
particular, the commutator of any left-invariant vector fields must be a linear
combination of the generators, \(\CT{\vec e_i}{\vec e_j} = c^k_{ij}\vec e_k\)
where the \(c^k_{ij}\in\R\) are called the \emph{structure constants}.  This
makes the linear space of left-invariant vector fields into a \emph{Lie
algebra}.

Any left-invariant vector field \(\vec v\) has an integral
curve \footnote{Strictly speaking this is only true locally: to be precise we
should write \(c:\I\to\M\) where \(\I\subseteq\R\) is a neighbourhood of zero.}
\(c:\R\to \G\) satisfying \(c(0)=\ident\).  Along this curve we have an Abelian
subgroup of \(\G\) satisfying \(c(s+t) = c(s)c(t)\), so it is naturally to call
\(c\) an \emph{exponential map}, and write it as \(c(t) = \exp(\vec vt)\).  If
we view this as a function of \(\vec v\) then this defines a \emph{local flow}
of \(\vec v\), and is a map from the Lie algebra into the Lie group, \(\exp:
T\G\to\G\).

The \emph{commutator} of two elements \(g,h\in\G\) is defined to be \(C(g.h)
\defn g^{-1}h^{-1}gh\); in a neighbourhood of the identity where \(g=\exp(\vec
ut)\) and \(h=\exp(\vec vt)\) we have
\begin{eqnarray*}
  \lefteqn{C(g,h) = \exp(-\vec ut)\exp(-\vec vt)\exp(\vec ut)\exp(\vec vt)} 
    \quad &&  \\
  &=& \left(\ident - \vec ut + \frac12(\vec ut)^2\right)
    \left(\ident - \vec vt + \frac12(\vec vt)^2\right) \\
  && \times \left(\ident + \vec ut + \frac12(\vec ut)^2\right)
    \left(\ident + \vec vt + \frac12(\vec vt)^2\right) + \O(t^3) \\
  &=& \ident + \CT{\vec u}{\vec v}t^2 + \O(t^3)
  = \exp(\CT{\vec u}{\vec v}t^2) + \O(t^3).
\end{eqnarray*}

\subsection{Maurer--Cartan Equations}

The commutation relations may be succinctly expressed in terms of the cotangent
space \(T^{*}\G\).  We introduce a set of left-invariant 1-forms
\(\form\theta^i\) (called a frame or {\it rep\`ere mobile}) dual to the
generators \(\form\theta^i(\vec e_j) = \delta^i_j\).  From \eqref{eq:d-1-form}
we have
\begin{eqnarray*}
  \d\form\theta^i(\vec e_j,\vec e_k)
    &=& \vec e_j\form\theta^i(\vec e_k) - \vec e_k\form\theta^i(\vec e_j)
      - \form\theta^i(\CT{\vec e_j}{\vec e_k}) \\
    &=& \vec e_j\delta^i_k - \vec e_k\delta^i_j
      - \form\theta^i(c^\ell_{jk}\vec e_\ell)
    = -c^\ell_{jk}\delta^i_\ell = -c^i_{jk},
\end{eqnarray*}
so expanding the 2-form \(\d\form\theta^i = \alpha^i_{mn} \form\theta^m\wedge
\form\theta^n\) in terms of the basis 2-forms \(\form\theta^m\wedge \form
\theta^n\) we have
\begin{eqnarray*}
  \d\form\theta(\vec e_j,\vec e_k)
  &=& \alpha^i_{mn} \form\theta^m\wedge\form\theta^n(\vec e_j,\vec e_k) \\
  &=& \alpha^i_{mn} \left\{\form\theta^m(\vec e_j) \form\theta^n(\vec e_k)
    - \form\theta^m(\vec e_k) \form\theta^n(\vec e_j)\right\} \\
  &=& \alpha^i_{mn} \{\delta^m_j\delta^n_k - \delta^m_k\delta^n_j\}
  = \alpha^i_{jk} - \alpha^i_{kj},  
\end{eqnarray*}
thus the left-invariant forms \(\form\theta^i\) satisfy the Maurer--Cartan
equations \(d\form\theta^i = -\half c^i_{jk}\form\theta^j \wedge
\form\theta^k\) everywhere.

\subsection{Adjoint Representation}

For any Lie algebra the \emph{adjoint representation} is defined by \(\ad(\vec
x)\vec y \defn \CT{\vec x}{\vec y}\).  This is a representation of the Lie
algebra because for any \(\vec z\)
\begin{eqnarray*}
  \CT{\ad(\vec x)}{\ad(\vec y)}\vec z 
  &=& \ad(\vec x)\ad(\vec y)\vec z - \ad(\vec y)\ad(\vec x)\vec z \\
  &=& \ad(\vec x)\CT{\vec y}{\vec z} - \ad(\vec y)\CT{\vec x}{\vec z} \\
  &=& \CT{\vec x}{\CT{\vec y}{\vec z}} - \CT{\vec y}{\CT{\vec x}{\vec z}}
  = \CT{\CT{\vec x}{\vec y}}{\vec z} \\
  &=& \ad({\CT{\vec x}{\vec y}}){\vec z}
\end{eqnarray*}
where we used the Jacobi identity in the penultimate step, and thus
\(\CT{\ad(\vec x)}{\ad(\vec y)} = \ad({\CT{\vec x}{\vec y}})\).  In terms of
basis vectors we have \(\ad(\vec e_i)\vec e_j = \CT{\vec e_i}{\vec e_j} =
c^k_{ij}\vec e_k\), giving the explicit matrices \(\ad(\vec e_i)^k_j =
c^k_{ij}\).

\subsection{Cartan--Killing Metric} \label{sec:cartan-killing}

We may use the adjoint representation to define the \emph{Cartan--Killing
metric} on the Lie algebra as a trace, \(\CK{\vec x}{\vec y} \defn \Tr
[\ad(\vec x) \ad(\vec y)]/C_A\) where \(C_A\) is a constant; in terms of the
basis vectors \(g_{ij} \defn \CK{\vec e_i}{\vec e_j} = \Tr[\ad(\vec e_i)
  \ad(\vec e_j)]/C_A = c^k_{i\ell}c^\ell_{jk}/C_A\).  For a \emph{semi-simple}
Lie algebra the Cartan--Killing metric is non-singular and has an inverse
satisfying \(g^{ij}g_{jk} = \delta^i_k\).  For a \emph{simple} Lie algebra the
adjoint representation is irreducible, so by Schur's lemma the invariant
Cartan--Killing metric is a multiple of the unit matrix; we shall choose the
constant \(C_A\) such that this multiple is unity.  For \(\su(N)\) where the
generators in the defining \(N\) dimensional fundamental representation \(T_i\)
satisfy the commutation relations \(\CT{T_i}{T_j} = c^k_{ij}T_k\) and are
normalized such that \(\Tr T_iT_j = a\delta_{ij}\) the Cartan--Killing metric
is explicitly \(g_{ij}=\delta_{ij}\) with \(C_A=2aN\).

For semi-simple Lie algebras we can use the Cartan--Killing metric and its
inverse to lower and raise indices at will, for example we shall define \(p^i
\defn g^{ij}p_i\), and correspondingly we have an invariant quadratic form for
1-forms, \(\CK{\form\alpha}{\form\beta} = g^{ij}\alpha_i\beta_j\) where
\(\form \alpha = \alpha_i\theta^i\) and \(\form\beta = \beta_i\theta^i\).  We
also note that the quantity \(c_{ijk} = g_{i\ell} c^\ell_{jk} = -c_{ikj}\) is
totally antisymmetric, because \(\CK{\CT{\vec e_i}{\vec e_j}}{\vec e_k} =
c^\ell_{ij} \CK{\vec e_\ell}{\vec e_k} = c^\ell_{ij} g_{\ell k} = c_{kij}\),~and
\begin{eqnarray*}
  \lefteqn{C_A\CK{\CT {\vec X}{\vec Y}}{\vec Z}
  = \Tr\Bigl(\ad(\CT {\vec X}{\vec Y})\ad({\vec Z})\Bigr)} \qquad && \\
  &=& \Tr\Bigl(\CT{\ad({\vec X})}{\ad({\vec Y})}\ad({\vec Z})\Bigr) \\
  &=& \Tr\Bigl(\ad({\vec X})\ad({\vec Y})\ad({\vec Z})
    - \ad({\vec Y})\ad({\vec X})\ad({\vec Z})\Bigr) \\
  &=& \Tr\Bigl(\ad({\vec Z})\ad({\vec X})\ad({\vec Y})
    - \ad({\vec X})\ad({\vec Z})\ad({\vec Y})\Bigr) \\
  &=& \Tr\Bigl(\CT{\ad({\vec Z})}{\ad({\vec X})}\ad({\vec Y})\Bigr) \\
  &=& \Tr\Bigl(\ad(\CT {\vec Z}{\vec X})\ad({\vec Y})\Bigr)
  = C_A\CK{\CT {\vec Z}{\vec X}}{\vec Y},
\end{eqnarray*}
hence \(c_{ijk} = c_{jki} = c_{kij}\).

\ifprd
\bibliographystyle{apsrev}
\fi

\begin{thebibliography}{29}
\expandafter\ifx\csname natexlab\endcsname\relax\def\natexlab#1{#1}\fi
\expandafter\ifx\csname bibnamefont\endcsname\relax
  \def\bibnamefont#1{#1}\fi
\expandafter\ifx\csname bibfnamefont\endcsname\relax
  \def\bibfnamefont#1{#1}\fi
\expandafter\ifx\csname citenamefont\endcsname\relax
  \def\citenamefont#1{#1}\fi
\expandafter\ifx\csname url\endcsname\relax
  \def\url#1{\texttt{#1}}\fi
\expandafter\ifx\csname urlprefix\endcsname\relax\def\urlprefix{URL }\fi
\providecommand{\bibinfo}[2]{#2}
\providecommand{\eprint}[2][]{\url{#2}}

\bibitem[{\citenamefont{Duane et~al.}(1987)\citenamefont{Duane, Kennedy,
  Pendleton, and Roweth}}]{duane87a}
\bibinfo{author}{\bibfnamefont{S.}~\bibnamefont{Duane}},
  \bibinfo{author}{\bibfnamefont{A.~D.} \bibnamefont{Kennedy}},
  \bibinfo{author}{\bibfnamefont{B.~J.} \bibnamefont{Pendleton}},
  \bibnamefont{and} \bibinfo{author}{\bibfnamefont{D.}~\bibnamefont{Roweth}},
  \bibinfo{journal}{Phys. Lett.} \textbf{\bibinfo{volume}{195B}},
  \bibinfo{pages}{216} (\bibinfo{year}{1987}).

\bibitem[{\citenamefont{Hairer et~al.}(2006)\citenamefont{Hairer, Lubich, and
  Wanner}}]{hairer:2006}
\bibinfo{author}{\bibfnamefont{E.}~\bibnamefont{Hairer}},
  \bibinfo{author}{\bibfnamefont{C.}~\bibnamefont{Lubich}}, \bibnamefont{and}
  \bibinfo{author}{\bibfnamefont{G.}~\bibnamefont{Wanner}},
  \emph{\bibinfo{title}{Geometric Numerical Integration}}
  (\bibinfo{publisher}{Springer}, \bibinfo{year}{2006}), ISBN
  \bibinfo{isbn}{364205157X}, \bibinfo{note}{second edition}.

\bibitem[{\citenamefont{Omelyan et~al.}(2001)\citenamefont{Omelyan, Mryglod,
  and Folk}}]{omelyan:2001}
\bibinfo{author}{\bibfnamefont{I.~P.} \bibnamefont{Omelyan}},
  \bibinfo{author}{\bibfnamefont{I.~M.} \bibnamefont{Mryglod}},
  \bibnamefont{and} \bibinfo{author}{\bibfnamefont{R.}~\bibnamefont{Folk}},
  \bibinfo{journal}{Phys. Rev. Lett.} \textbf{\bibinfo{volume}{86}},
  \bibinfo{pages}{898} (\bibinfo{year}{2001}).

\bibitem[{\citenamefont{Takaishi}(2000)}]{Takaishi:1999bi}
\bibinfo{author}{\bibfnamefont{T.}~\bibnamefont{Takaishi}},
  \bibinfo{journal}{Comput. Phys. Commun.} \textbf{\bibinfo{volume}{133}},
  \bibinfo{pages}{6} (\bibinfo{year}{2000}), \eprint{hep-lat/9909134}.

\bibitem[{\citenamefont{Takaishi and de~Forcrand}(2006)}]{forcrand:2006}
\bibinfo{author}{\bibfnamefont{T.}~\bibnamefont{Takaishi}} \bibnamefont{and}
  \bibinfo{author}{\bibfnamefont{P.}~\bibnamefont{de~Forcrand}},
  \bibinfo{journal}{Phys. Rev. E} \textbf{\bibinfo{volume}{73}},
  \bibinfo{pages}{036706} (\bibinfo{year}{2006}).

\bibitem[{\citenamefont{Drummond et~al.}(1983)\citenamefont{Drummond, Duane,
  and Horgan}}]{drummond83a}
\bibinfo{author}{\bibfnamefont{I.~T.} \bibnamefont{Drummond}},
  \bibinfo{author}{\bibfnamefont{S.}~\bibnamefont{Duane}}, \bibnamefont{and}
  \bibinfo{author}{\bibfnamefont{R.~R.} \bibnamefont{Horgan}},
  \bibinfo{journal}{Nucl. Phys.} \textbf{\bibinfo{volume}{B220 [FS8]}},
  \bibinfo{pages}{119} (\bibinfo{year}{1983}).

\bibitem[{\citenamefont{Kennedy and Rossi}(1989)}]{kennedy88b}
\bibinfo{author}{\bibfnamefont{A.~D.} \bibnamefont{Kennedy}} \bibnamefont{and}
  \bibinfo{author}{\bibfnamefont{P.}~\bibnamefont{Rossi}},
  \bibinfo{journal}{Nucl. Phys.} \textbf{\bibinfo{volume}{B327}},
  \bibinfo{pages}{782} (\bibinfo{year}{1989}).

\bibitem[{\citenamefont{Bishop and Goldberg}(1980)}]{bishop:1980}
\bibinfo{author}{\bibfnamefont{R.~L.} \bibnamefont{Bishop}} \bibnamefont{and}
  \bibinfo{author}{\bibfnamefont{S.~I.} \bibnamefont{Goldberg}},
  \emph{\bibinfo{title}{Tensor Analysis on Manifolds}}
  (\bibinfo{publisher}{Dover}, \bibinfo{year}{1980}), ISBN
  \bibinfo{isbn}{0-486-64039-6}.

\bibitem[{\citenamefont{Choquet-Bruhat
  et~al.}(1977)\citenamefont{Choquet-Bruhat, DeWitt-Morette, and
  Dillard-Bleick}}]{bruhat:1977}
\bibinfo{editor}{\bibfnamefont{Y.}~\bibnamefont{Choquet-Bruhat}},
  \bibinfo{editor}{\bibfnamefont{C.}~\bibnamefont{DeWitt-Morette}},
  \bibnamefont{and}
  \bibinfo{editor}{\bibfnamefont{M.}~\bibnamefont{Dillard-Bleick}}, eds.,
  \emph{\bibinfo{title}{Analysis, Manifolds and Physics}}
  (\bibinfo{publisher}{North-Holland}, \bibinfo{year}{1977}), ISBN
  \bibinfo{isbn}{0-7204-0494-0}.

\bibitem[{\citenamefont{Helgason}(1978)}]{helgason:1978}
\bibinfo{author}{\bibfnamefont{S.}~\bibnamefont{Helgason}},
  \emph{\bibinfo{title}{Differential Geometry, Lie Groups, and Symmetric
  Spaces}} (\bibinfo{publisher}{Academic Press}, \bibinfo{year}{1978}), ISBN
  \bibinfo{isbn}{0-12-338460-5}.

\bibitem[{\citenamefont{Hicks}(1971)}]{hicks:1971}
\bibinfo{author}{\bibfnamefont{N.~J.} \bibnamefont{Hicks}},
  \emph{\bibinfo{title}{Notes on Differential Geometry}}
  (\bibinfo{publisher}{van Nostrand Reinhold}, \bibinfo{year}{1971}), ISBN
  \bibinfo{isbn}{0-442-034059}.

\bibitem[{\citenamefont{Spivak}(1970)}]{spivak:1970}
\bibinfo{author}{\bibfnamefont{M.}~\bibnamefont{Spivak}},
  \emph{\bibinfo{title}{Differential Geometry}} (\bibinfo{publisher}{Publish or
  Perish}, \bibinfo{year}{1970}), ISBN \bibinfo{isbn}{0-914098-00-4}.

\bibitem[{\citenamefont{Omelyan et~al.}(2002)\citenamefont{Omelyan, Mryglod,
  and Folk}}]{omelyan:2002}
\bibinfo{author}{\bibfnamefont{I.~P.} \bibnamefont{Omelyan}},
  \bibinfo{author}{\bibfnamefont{I.~M.} \bibnamefont{Mryglod}},
  \bibnamefont{and} \bibinfo{author}{\bibfnamefont{R.}~\bibnamefont{Folk}},
  \bibinfo{journal}{Phys. Rev. E} \textbf{\bibinfo{volume}{66}},
  \bibinfo{pages}{026701} (\bibinfo{year}{2002}).

\bibitem[{\citenamefont{Omelyan et~al.}(2003)\citenamefont{Omelyan, Mryglod,
  and Folk}}]{omelyan:2003}
\bibinfo{author}{\bibfnamefont{I.~P.} \bibnamefont{Omelyan}},
  \bibinfo{author}{\bibfnamefont{I.~M.} \bibnamefont{Mryglod}},
  \bibnamefont{and} \bibinfo{author}{\bibfnamefont{R.}~\bibnamefont{Folk}},
  \bibinfo{journal}{Comp. Phys. Commun.} \textbf{\bibinfo{volume}{151}},
  \bibinfo{pages}{273} (\bibinfo{year}{2003}).

\bibitem[{\citenamefont{Clark et~al.}(2008)\citenamefont{Clark, Kennedy, and
  Silva}}]{Clark:2008gh}
\bibinfo{author}{\bibfnamefont{M.~A.} \bibnamefont{Clark}},
  \bibinfo{author}{\bibfnamefont{A.~D.} \bibnamefont{Kennedy}},
  \bibnamefont{and} \bibinfo{author}{\bibfnamefont{P.~J.} \bibnamefont{Silva}},
  \bibinfo{journal}{PoS} \textbf{\bibinfo{volume}{LAT2008}},
  \bibinfo{pages}{041} (\bibinfo{year}{2008}), \eprint{0810.1315}.

\bibitem[{\citenamefont{Clark et~al.}(2010)\citenamefont{Clark, {Jo\'o},
  Kennedy, and Silva}}]{Clark:2010qw}
\bibinfo{author}{\bibfnamefont{M.~A.} \bibnamefont{Clark}},
  \bibinfo{author}{\bibfnamefont{B.}~\bibnamefont{{Jo\'o}}},
  \bibinfo{author}{\bibfnamefont{A.~D.} \bibnamefont{Kennedy}},
  \bibnamefont{and} \bibinfo{author}{\bibfnamefont{P.~J.} \bibnamefont{Silva}},
  \bibinfo{journal}{PoS} \textbf{\bibinfo{volume}{LATTICE2010}},
  \bibinfo{pages}{323} (\bibinfo{year}{2010}), \eprint{1011.0230}.

\bibitem[{\citenamefont{Clark et~al.}(2011)\citenamefont{Clark, {Jo\'o},
  Kennedy, and Silva}}]{Clark:2011ir}
\bibinfo{author}{\bibfnamefont{M.~A.} \bibnamefont{Clark}},
  \bibinfo{author}{\bibfnamefont{B.}~\bibnamefont{{Jo\'o}}},
  \bibinfo{author}{\bibfnamefont{A.~D.} \bibnamefont{Kennedy}},
  \bibnamefont{and} \bibinfo{author}{\bibfnamefont{P.~J.} \bibnamefont{Silva}},
  \bibinfo{journal}{Phys. Rev.} \textbf{\bibinfo{volume}{D84}},
  \bibinfo{pages}{071502} (\bibinfo{year}{2011}), \eprint{1108.1828}.

\bibitem[{\citenamefont{Kennedy et~al.}(2009)\citenamefont{Kennedy, Clark, and
  Silva}}]{Kennedy:2009fe}
\bibinfo{author}{\bibfnamefont{A.~D.} \bibnamefont{Kennedy}},
  \bibinfo{author}{\bibfnamefont{M.~A.} \bibnamefont{Clark}}, \bibnamefont{and}
  \bibinfo{author}{\bibfnamefont{P.~J.} \bibnamefont{Silva}},
  \bibinfo{journal}{PoS} \textbf{\bibinfo{volume}{LAT2009}},
  \bibinfo{pages}{021} (\bibinfo{year}{2009}), \eprint{0910.2950}.

\bibitem[{\citenamefont{Morrin et~al.}(2006)\citenamefont{Morrin, {O'Cais},
  Peardon, Ryan, and Skullerud}}]{Morrin:2006tf}
\bibinfo{author}{\bibfnamefont{R.}~\bibnamefont{Morrin}},
  \bibinfo{author}{\bibfnamefont{A.}~\bibnamefont{{O'Cais}}},
  \bibinfo{author}{\bibfnamefont{M.}~\bibnamefont{Peardon}},
  \bibinfo{author}{\bibfnamefont{S.~M.} \bibnamefont{Ryan}}, \bibnamefont{and}
  \bibinfo{author}{\bibfnamefont{J.-I.} \bibnamefont{Skullerud}},
  \bibinfo{journal}{Phys. Rev.} \textbf{\bibinfo{volume}{D74}},
  \bibinfo{pages}{014505} (\bibinfo{year}{2006}), \eprint{hep-lat/0604021}.

\bibitem[{\citenamefont{Lin et~al.}(2009)}]{Lin:2008pr}
\bibinfo{author}{\bibfnamefont{H.-W.} \bibnamefont{Lin}} \bibnamefont{et~al.}
  (\bibinfo{collaboration}{{Hadron} {Spectrum} {Collaboration}}),
  \bibinfo{journal}{Phys. Rev.} \textbf{\bibinfo{volume}{D79}},
  \bibinfo{pages}{034502} (\bibinfo{year}{2009}), \eprint{0810.3588}.

\bibitem[{\citenamefont{Hasenbusch}(2001)}]{Hasenbusch:2001ne}
\bibinfo{author}{\bibfnamefont{M.}~\bibnamefont{Hasenbusch}},
  \bibinfo{journal}{Phys. Lett.} \textbf{\bibinfo{volume}{B519}},
  \bibinfo{pages}{177} (\bibinfo{year}{2001}), \eprint{hep-lat/0107019}.

\bibitem[{\citenamefont{Hasenbusch and Jansen}(2003)}]{Hasenbusch:2002ai}
\bibinfo{author}{\bibfnamefont{M.}~\bibnamefont{Hasenbusch}} \bibnamefont{and}
  \bibinfo{author}{\bibfnamefont{K.}~\bibnamefont{Jansen}},
  \bibinfo{journal}{Nucl. Phys.} \textbf{\bibinfo{volume}{B659}},
  \bibinfo{pages}{299} (\bibinfo{year}{2003}), \eprint{hep-lat/0211042}.

\bibitem[{\citenamefont{{L\"uscher}}(2005)}]{Luscher:2005rx}
\bibinfo{author}{\bibfnamefont{M.}~\bibnamefont{{L\"uscher}}},
  \bibinfo{journal}{Comput. Phys. Commun.} \textbf{\bibinfo{volume}{165}},
  \bibinfo{pages}{199} (\bibinfo{year}{2005}), \eprint{hep-lat/0409106}.

\bibitem[{\citenamefont{Clark and Kennedy}(2007{\natexlab{a}})}]{Clark:2006fx}
\bibinfo{author}{\bibfnamefont{M.~A.} \bibnamefont{Clark}} \bibnamefont{and}
  \bibinfo{author}{\bibfnamefont{A.~D.} \bibnamefont{Kennedy}},
  \bibinfo{journal}{Phys. Rev. Lett.} \textbf{\bibinfo{volume}{98}},
  \bibinfo{pages}{051601} (\bibinfo{year}{2007}{\natexlab{a}}),
  \eprint{hep-lat/0608015}.

\bibitem[{\citenamefont{Clark and Kennedy}(2007{\natexlab{b}})}]{Clark:2006wp}
\bibinfo{author}{\bibfnamefont{M.~A.} \bibnamefont{Clark}} \bibnamefont{and}
  \bibinfo{author}{\bibfnamefont{A.~D.} \bibnamefont{Kennedy}},
  \bibinfo{journal}{Phys. Rev.} \textbf{\bibinfo{volume}{D75}},
  \bibinfo{pages}{011502} (\bibinfo{year}{2007}{\natexlab{b}}),
  \eprint{hep-lat/0610047}.

\bibitem[{\citenamefont{Berndt}(2001)}]{berndt:2001}
\bibinfo{author}{\bibfnamefont{R.}~\bibnamefont{Berndt}},
  \emph{\bibinfo{title}{An Introduction to Symplectic Geometry}},
  vol.~\bibinfo{volume}{21} of \emph{\bibinfo{series}{Graduate Studies in
  Mathematics}} (\bibinfo{publisher}{American Mathematical Society},
  \bibinfo{year}{2001}), ISBN \bibinfo{isbn}{0-8218-2056-7}.

\bibitem[{\citenamefont{Clark and Kennedy}(2007{\natexlab{c}})}]{Clark:2007ff}
\bibinfo{author}{\bibfnamefont{M.~A.} \bibnamefont{Clark}} \bibnamefont{and}
  \bibinfo{author}{\bibfnamefont{A.~D.} \bibnamefont{Kennedy}},
  \bibinfo{journal}{Phys. Rev.} \textbf{\bibinfo{volume}{D76}},
  \bibinfo{pages}{074508} (\bibinfo{year}{2007}{\natexlab{c}}),
  \eprint{0705.2014}.

\bibitem[{\citenamefont{Gottlieb et~al.}(1987)\citenamefont{Gottlieb, Liu,
  Toussaint, Renken, and Sugar}}]{gottlieb87a}
\bibinfo{author}{\bibfnamefont{S.}~\bibnamefont{Gottlieb}},
  \bibinfo{author}{\bibfnamefont{W.}~\bibnamefont{Liu}},
  \bibinfo{author}{\bibfnamefont{D.}~\bibnamefont{Toussaint}},
  \bibinfo{author}{\bibfnamefont{R.~L.} \bibnamefont{Renken}},
  \bibnamefont{and} \bibinfo{author}{\bibfnamefont{R.~L.} \bibnamefont{Sugar}},
  \bibinfo{journal}{Phys. Rev.} \textbf{\bibinfo{volume}{D35}},
  \bibinfo{pages}{2531} (\bibinfo{year}{1987}).

\bibitem[{\citenamefont{Yin and Mawhinney}(2011)}]{Yin:2011sz}
\bibinfo{author}{\bibfnamefont{H.}~\bibnamefont{Yin}} \bibnamefont{and}
  \bibinfo{author}{\bibfnamefont{R.~D.} \bibnamefont{Mawhinney}},
  \bibinfo{journal}{PoS} \textbf{\bibinfo{volume}{LATTICE2011}},
  \bibinfo{pages}{051} (\bibinfo{year}{2011}), \eprint{1111.5059}.

\end{thebibliography}

\end{document}